\documentclass[twoside,onecolumn]{article}

\usepackage{tabulary,graphicx,times,caption,fancyhdr,amsfonts,amssymb,amsbsy,latexsym,amsmath}
\usepackage[utf8]{inputenc}
\usepackage{url,multirow,morefloats,floatflt,cancel,tfrupee,textcomp,colortbl,xcolor,pifont}
\usepackage[nointegrals]{wasysym}
\usepackage{todonotes}
\makeatletter

\usepackage{ifxetex}
\ifxetex\else
  \usepackage{dblfloatfix}
\fi

\@ifundefined{subparagraph}{
\def\subparagraph{\@startsection{paragraph}{5}{2\parindent}{0ex plus 0.1ex minus 0.1ex}%
{0ex}{\normalfont\small\itshape}}%
}{}

\def\URL#1#2{\@ifundefined{href}{#2}{\href{#1}{#2}}}

\def\UrlOrds{\do\*\do\-\do\~\do\'\do\"\do\-}%
\g@addto@macro{\UrlBreaks}{\UrlOrds}

\makeatother

\usepackage[paperheight=11in,paperwidth=8.3in,margin=2.5cm,headsep=.7cm,top=2.5cm]{geometry}
\usepackage[T1]{fontenc}

\renewenvironment{abstract}
	{\trivlist\item[]\leftskip0pt\par\vskip4pt\noindent
  	\textbf{\abstractname}\mbox{\null}\\}
	{\par\noindent\endtrivlist}

\def\keywords#1{\par\medskip\par\noindent\textbf{Keywords}: #1\par}

 \date{} \emergencystretch 8pt

\makeatletter
\def\author#1{\gdef\@author{\hskip-\tabcolsep%
	\parbox{\textwidth}{\raggedright\bfseries#1\\[1pc]}}}
\def\address[#1]#2{\g@addto@macro\@author{\\\hskip-\tabcolsep\parbox{\textwidth}{\raggedright%
	\normalsize\normalfont\textsuperscript{#1}#2}}}
\let\addresslink\textsuperscript
\def\correspondence#1{\g@addto@macro\@author{\\\hskip-\tabcolsep\parbox{\textwidth}{\raggedright%
	\vspace*{10pt}\normalsize\normalfont~\\#1~\\[12pt]}}}
\def\email#1{\g@addto@macro\@author{\\\hskip-\tabcolsep\parbox{\textwidth}{\raggedright%
	\normalsize\normalfont Emails: #1}}}

\def\title#1{\gdef\@title{\vspace*{-30pt}%
	\raggedright\textbf{\@journaltitle}~\\%
  \raggedright\bfseries\ifx\@articleType\@empty\vspace*{20pt}\else%
  \vspace*{20pt}\@articleType\vspace*{20pt}\\\fi#1}}
\let\@journaltitle\@empty \def\journaltitle#1{\gdef\@journaltitle{{\normalfont\itshape#1}}}
\let\@articleType\@empty \def\articletype#1{\gdef\@articleType{{\normalfont\itshape#1}}}

\let\@runningHead\@empty \def\RunningHead#1{\gdef\@runningHead{{\normalfont #1}}}

\usepackage{fancyhdr}
\fancypagestyle{headings}{\fancyhf{}
  \fancyhead[R]{\itshape\@runningHead}
  \fancyfoot[C]{\thepage}}
\fancypagestyle{plain}{%
	\fancyhf{}\fancyhead[R]{\LaTeX\ template for Hindawi journal articles}
  \fancyfoot[C]{\thepage}}
\makeatother

\usepackage[%
	numbers,sort&compress%
  ]{natbib}

\usepackage{float,xcolor}

\journaltitle{Conference Issue: Interdisciplinary and Sustainable Cybersecurity}
\begin{document}

\title{Towards an Improved Taxonomy of Attacks related to Digital Identities and Identity Management Systems}

\author{%
		Daniela Pöhn\addresslink{1} and
  	Wolfgang Hommel\addresslink{1}
    }

\address[1]{University of Bundeswehr Munich, RI CODE, 85577 Neubiberg, Germany}

\correspondence{Correspondence should be addressed to
    	Daniela Pöhn: daniela.poehn@unibw.de}


\RunningHead{Running head}

\maketitle

\begin{abstract}
Digital transformation with the adoption of cloud technologies, outsourcing, and working-from-home possibilities permits flexibility for organizations and persons. At the same time, it makes it more difficult to secure the IT infrastructure as the IT team needs to keep track of who is accessing what data from where and when on which device. With these changes, identity management as a key element of security becomes more important. Identity management relates to the technologies and policies for the identification, authentication, and authorization of users (humans, devices) in computer networks. Due to the diversity of identity management (i.\,e., models, protocols, and implementations), different requirements, problems, and attack vectors need to be taken into account. In order to secure identity management systems with their identities, a systematic approach is required. In this article, we propose the improved framework Taxonomy for Identity Management related to Attacks (TaxIdMA). The purpose of TaxIdMA is to classify existing attacks, attack vectors, and vulnerabilities associated with system identities, identity management systems, and end-user identities. In addition, the background of these attacks can be described in a structured and systematic way. The taxonomy is applied to the Internet of Things and self-sovereign identities. It is enhanced by a description language for threat intelligence sharing. Last but not least,  TaxIdMA is evaluated and improved based on expert interviews, statistics, and discussions. This step enables broader applicability and level of detail at the same time. The combination of TaxIdMA, which allows a structured way to outline attacks and is applicable to different scenarios, and a description language for threat intelligence help to improve the security identity management systems and processes.
\end{abstract}

\section{Introduction}

Credential theft and social engineering are the most frequent attacks that organizations are facing~\cite{ponemon}. With valid credentials, miscreants can start their attacks on organizations or sell them for financial purposes. The difficulties often begin with password management: many accounts require secure passwords. While a password manager helps to generate and remember passwords, not all persons use them effectively or at all~\cite{263832,238303,281310}. Therefore, weak passwords such as \texttt{123456}, \texttt{qwertz}, and \texttt{password} are common. These are included in wordlists with \texttt{rockyou.txt}~\cite{rockyou} (14,341,564 unique passwords used in 32,603,388 accounts) being the best known. The password list of john~\cite{john} is typically applied during cracking, while brutespray~\cite{brutespray} with its wordlist automates brute-forcing based on the Network Mapper (nmap) output. In consequence, attackers can easily break them. Even if the users apply passwords with high entropy, they might be reused for several platforms~\cite{10.1145/3183341}, enabling credential stuffing attacks if one such system is compromised or the password is stolen otherwise~\cite{10.1145/3460120.3484791}. In addition, social engineering does not target the storage or complexity of passwords but the human element. This shows that credentials and, thereby, identity management (IdM) are core elements for security in a network.

While these user identities seem to be an easy target, identity management systems (IdMS), which store and manage the identities of an organization, can be targeted in more serious attacks as shown by the SolarWinds incident~\cite{10.1145/3449047,9382367,9604375}. With access to an identity management system, which serves as a central identity repository, attackers also have access to any resource (service, computer, printer, etc.)~\cite{mci/Fritsch2020}. The consequences are complex, ranging from lost data and compromised accounts to financial loss. As a result, identity management systems need to be secured and well-protected. Typical defense mechanisms are strong password policies, usage of password managers, enforcement of multi-factor authentication, privileged access management, and training. Depending on the implemented identity management system, specific defense mechanisms and configurations may be applied. According to the Purple Knight Report 2022~\cite{purpleknight}, organizations have problems correctly securing Microsoft Active Directory (AD) as one of the most prominent identity management systems.

In order to systematically analyze and evaluate attacks, attack vectors, and vulnerabilities, a structured approach is required. In this context, taxonomies provide an overview of these complex systems and, thereby, possible attacks. Such a systematic approach helps to enhance the current situation by identifying gaps and providing guidelines for new security mechanisms. To the best of our knowledge, although several taxonomies and categorizations have been proposed and some case studies on specific attacks have been published, only our previous work~\cite{10.1145/3538969.3544430} is a taxonomy on attacks targeting identity management. In consequence, we propose an improved taxonomy framework for attacks related to identities and identity management systems, short TaxIdMA. This framework consists of 1) a background description, 2) taxonomies for attacks on end-user identities, system identities, and identity management systems, and 3) application on Internet of Things (IoT) identities and self-sovereign identities (SSI). In order to exchange information about these attacks, an extension to the description language Structured Threat Information Expression (STIX)~\cite{stix} is proposed.

This article extends~\cite{10.1145/3538969.3544430} by a background description, improved TaxIdMA taxonomies, the addition of IoT and SSI, as well as an enhanced evaluation of the related work and TaxIdMA. Last but not least, an extension of the description language STIX for exchanging attack information related to identity management is proposed. As a result, the contribution of the article is multi-fold: 1) an improved taxonomy framework for attacks on identities and identity management systems;  2) an extended evaluation of related work; 3) an extended evaluation of TaxIdMA; 4) an extension of the description language STIX based on TaxIdMA.

The remainder of this article is as follows: First, the background of identity management is summarized, followed by a broad discussion of various aspects of related work. Then, the methodology to establish and verify the taxonomy framework is outlined. The methodology includes information about the changes from the previous to the current version of TaxIdMA. This is followed by TaxIdMA, the taxonomy framework for attacks with background, attacks on end-user, system identities, and identity management systems. TaxIdMA is then applied to the areas of IoT and SSI. This enhanced version of TaxIdMA is evaluated by expert interviews and statistical information.  Based on TaxIdMA, an extension of the description language STIX to exchange information related to identity management is proposed. Both aspects, TaxIdMA and description language, are discussed in the following section. Last but not least, a summary and an outlook on future work are given.

\section{Background}
\label{sec:basics}

Identity management is the organizational and technical process for registering and authorizing access rights during enrollment and authentication as well as controlling identities based on previously authorized access rights. Identity management is described by accessing a service from the user's perspective, followed by centralized, federated, and user-centric identity management, and concluded with issues related to identity management.

\subsection{End-Users}

In order to access different services, users first need to register. Users typically add personal information, the so-called attributes. Then, users can authenticate and get authorized to access the requested services. Authentication is often password-based, though other authentication methods from the categories of knowledge, possession, and biometrics can be applied as well. With an increasing number of services, users tend to forget their credentials. This frequently results in the reuse of passwords across several accounts, which reduces security. In contrast, users could use password managers to store and generate passwords for multiple accounts. To improve security, multi-factor authentication (MFA)~\cite{10.1145/3394788.3394789} might be implemented. This means that at least two different and independent methods are required. A more usable version is risk-based authentication (RBA)~\cite{10.1145/3546069}, where additional factors are requested depending on risk and the user's situation. If one authentication method is (temporarily) unavailable, predefined fallback mechanisms, such as security questions and email links, come into play~\cite{10.1145/2785830.2785839}.

\subsection{Local Identity Management}

Windows accounts can be categorized as the user, administrator, and system. Administrator accounts generally have higher permissions than user accounts. They have full control of the files, directories, services, and other resources on the local computer. At the same time, they are able to manage accounts, rights, and permissions. On the other hand, users are more restricted in their rights. Depending on roles among other things, their permissions vary. In an illicitly configured network, a user could theoretically have more or less the same permissions as an administrator~\cite{10.1145/1837110.1837112}. A similar system exists for Linux, where the highest privileges come with the user type root. When installing services on Linux, they typically come with a corresponding user. For example, web services and PHP often use \texttt{www-data} in the group \texttt{www-data}. Those service users have limited permissions due to security reasons.

A pluggable authentication module (PAM)~\cite{10.1145/238168.238177} is a mechanism to integrate multiple authentication possibilities. Thereby, PAM allows programs to reuse these schemes instead of requiring each developer to write their own method. For example, Linux PAM is a suite of libraries that allow system administrators to configure authentication methods, such as local passwords and lightweight directory access protocol (LDAP), for their users. PAM can be used to streamline local identity management.

\subsection{Centralized Identity Management}

Identity management allows users to access different services. Thereby, the security of the services is tightly tied to it. In organizations, typically an identity management system is operated. The introduction of LDAP~\cite{4769598} started the first evolution of identity management towards a central system. LDAP maintains and shares distributes directory information services, such as users, networks, services, and applications. Popular implementations include Open\-LDAP and Microsoft AD, which combines LDAP with the Kerberos protocol. Typical issues with AD~\cite{9953407} can be boiled down to privileged account activity, login failures, and remote logins. As shown in the introduction, properly securing AD is though not an easy task. With single sign-on (SSO) running, the user has to log in only once to access several services. Identity management can also be operated on cloud services~\cite{Kostopoulos:2017:TAS:3098954.3104061}.

\subsection{Federated Identity Management}

As organizations tend to cooperate, there are two main possibilities related to identity management: 1) duplicate the accounts and 2) implement federated identity management (FIM). FIM~\cite{4489845} allows users to utilize their home organizations' credentials to sign in to other services within the trust boundaries of the federation. It thereby uses centralized identity management as the main source. In consequence, users only have to remember the credentials for their accounts at their home organizations. At the same time, security incidents may have a bigger impact. Two main FIM directions are used in practice: 1) Security Assertion Markup Language (SAML)~\cite{samloverview} and 2) Open Authorization (OAuth) 2.0~\cite{RFC6749} for authorization with OpenID Connect (OIDC)~\cite{openidconnect} for authentication. OAuth and OIDC can be combined but may run on their own. Typical use cases include research and education and eID federations~\cite{info10060210} (SAML) resp. commercial web services (OAuth and OIDC). Both directions with the protocols, implementations, and configurations have different security issues though~\cite{7961984,oidcsec,10.1145/2976749.2978385,I-D.ietf-oauth-security-topics,RFC6819,samlsec}.

\subsection{User-Centric Identity Management}

In parallel, several user-centric identity management approaches were proposed and introduced, in order to give users more control over their data. One protocol is User Managed Access (UMA)~\cite{umagrant,umafed,10.1145/1866855.1866865}, which is built upon OAuth. With SSI~\cite{8776589,8713271}, the research direction gained momentum. In contrast to traditional identity management, the user has full control over their data, which is issued by issuers, formerly home organizations, in form of verifiable credentials. These are stored in the user's wallet and can be rearranged as verifiable presentations to holders, which offer services. Further information is stored decentralized, for example by the use of blockchain. So far, Naik et al.~\cite{9659929} published the only approach to systematically evaluate the security of SSI.  Although this is a systematic approach, further attack vectors are possible, such as obtaining administrator credentials by leaks, configuration issues, etc.

\subsection{Issues with Identity Management}

As shown by the user side, not only do organizations operate identity management for their employees and cooperation, but also for different purposes, such as customer identity management.  For web services, the identity data is often stored in databases, but also other forms of management are possible. Identity management for IoT devices might be used in parallel, leading to a multitude of identity management systems within the identity models~\cite{7804984}. This makes it harder to configure and secure the systems correctly.

\subsection{Summary}

When we look at the aforementioned areas of identity management, it becomes clear that involved entities, protocols with their implementations, technical requirements, and managed identities are different. For this reason, the improved taxonomy must allow the inclusion of all the heterogeneous aspects to categorize attacks systematically without excluding relevant information.

\section{Related Work}
\label{sec:sota}

In this section, related work towards attack categories and generic as well as specific attack taxonomies is discussed. To find the generic approaches, we used the search terms \textit{(taxonomy OR categorization OR classification) AND (attack OR threat OR security)}. For the specific attack taxonomies, we added \textit{AND identity}. We applied these search terms at ACM, IEEE, Springer Link, USENIX, and MDPI. We excluded posters and short papers as well as publications, which apply taxonomies. This is enhanced by a description of related work on attacks on identity management and specific application. Here, we used the search terms \textit{(threat OR attack) AND (identity OR "internet of things" OR "self-sovereign identity")}. In addition, threat information-sharing approaches for both, languages and platforms, are evaluated. The corresponding search term is \textit{"threat information" AND sharing}. Last but not least, the limitations of related work are described.

\subsection{Attack Categories}

Several community-driven and commercial approaches categorize and list attacks~\cite{10.1145/3375408.3375410}. For example, the Common Weakness Enumeration (CWE) by MITRE~\cite{cwe} is a community-driven list of software and hardware weaknesses, which is applied as a common language for weakness identification, mitigation, and prevention. This includes weaknesses in identity management, such as different types of improper access controls ranging from Hypertext Transfer Protocol (HTTP) cookies to on-chip hardware issues. MITRE ATT\&CK~\cite{strom2018mitre} details numerous attack methods during the cyber kill chain, also called the attack lifecycle~\cite{8551383}. For example, during reconnaissance information is gathered through phishing and searches. The initial access includes phishing and the usage of valid accounts. Thereby, several identity-related methods are matched to the lifecycle. The Common Attack Pattern Enumerations and Classifications (CAPEC)~\cite{capec} describes attack patterns based on software design patterns. In the area of identity management, the subvert of access control comprises, for example, authentication abuse and bypass as well as physical theft. Even though these design patterns are important, not all attacks are based on them. The Open Web Application Security Project (OWASP)~\cite{owasp} publishes several top 10 lists and cheat sheets for typical problems, including error messages during login among others. Although these guidelines are relevant during configuration and improvements, they do not comprise all areas of identity management. Nonetheless, several aspects of these attack categories are taken into consideration while designing TaxIdMA.

\subsection{Attack Taxonomies}

Taxonomies are categorizations or classifications in mostly hierarchical order. Items are thereby arranged in groups or types. This can be used to organize and index knowledge. Originally from biology, the categorization is applied in different fields, including computer science.

\subsubsection{Generic Attack Taxonomies}

Different taxonomies related to attacks were proposed so far. Igure and Williams~\cite{4483667} analyze a multitude of taxonomies published from 1974 until 2006. Based on their findings, the authors propose a taxonomy of attacks and vulnerabilities in computer systems. Although the authors include numerous approaches, the resulting taxonomy is a generic attack taxonomy. As a consequence, it is not focused on a specific area. Chapman et al.~\cite{10.5555/2048558.2048569} propose a 3-tier taxonomy, which describes the effects of cyber attacks. The authors thereby report stages ranging from no access over user access to root access. In consequence, they specify different levels of permissions an attacker can gain. These levels may depend on the operating system (OS). In addition, the approach is unclear in some cases, such as service users like \texttt{www-data}. Derbyshire et al.~\cite{8406575} evaluate different well-known taxonomies based on pre-defined criteria and selected real-world attacks. The authors conclude that CAPEC outperforms other taxonomies, although several taxonomies do not include humans. As a result, TaxIdMA needs to include human elements. Cho et al.~\cite{8551383} explore different cyber kill chain models. Based on their evaluation, the authors propose a new model. Haber and Rolls summarize typical identity attack vectors in practice~\cite{vectors}. The cyber kill chain is a structured way to explain the stage of an attack and, therefore, should be included in the TaxIdMA.

\subsubsection{Specific Attack Taxonomies}

Other authors focus their taxonomies on a specific aspect. Habiba et al.~\cite{Habiba2014} propose a taxonomy related to cloud IdMS security issues. The authors first out the different identity management systems, before describing related security challenges and known attacks. These include brute-force attacks, cookie-replay attacks, data tampering attacks, eavesdropping, the elevation of privilege, identity theft, and phishing attacks. Although these attacks are relevant for identity management (identity management systems and end-users), they are limited to the at that time known attacks on cloud IdMS. Klaper and Hovy~\cite{10.1145/2612733.2612759} establish a taxonomy with cybersecurity topics. These basic topics are then linked to relevant educational or research material. Thereby, the authors notice gaps in the study curriculum. Different description languages are used to categorize and exchange threat information. Based on a literature review, Burger et al.~\cite{10.1145/2663876.2663883} propose a taxonomy related to the exchange of cyber threat intelligence information. The authors use a layered model with the 5 W's, intelligence, indicators, session, and transport. Within the category of the session, the authors differentiate authentication, authorization, and permissions. Although many aspects are relevant for attacks on identity management, the proposed taxonomy is rather generic. Husseis et al.~\cite{8888436} regard potential threats affecting biometric systems, while Mamchenko and Sabanov~\cite{8910969} explore USB-based attacks. Hollick et al.~\cite{10.1145/3041027.3041033} describe a taxonomy and attacker model for secure routing protocols. Chaipa et al.~\cite{9845581} propose a taxonomy of insider threats. These taxonomies are focused on a specific aspect and, therefore, are not suitable for identity management. Nonetheless, useful aspects, such as attack types, are repurposed for our taxonomy framework.

\subsubsection{IoT Attack Taxonomies}

Several IoT attack taxonomies are published. Alsamani and Lahza~\cite{8356843} propose an IoT taxonomy concerning security and privacy threats. The taxonomy consists of three dimensions, not covering the whole aspect. Nawir et al.~\cite{7804660} also present a security taxonomy, which includes some categories, but is already outdated. Khanam et al.~\cite{9256294} outline several security challenges. Based on the review, the authors design an attack taxonomy and corresponding countermeasures. The authors thereby use the layers application, network, physical, and multi. Although several attacks are outlined, the taxonomy is rather simple. Neshenko et al.~\cite{8688434} describe an extensive survey on IoT vulnerabilities. The authors differentiate the layers of devices, software, and network. Similarly, Wüstrich et al.~\cite{9219584} propose a simple naming scheme for IoT threat, which is used for a work-in-progress taxonomy. Rizvi et al.~\cite{8455902} differentiate architecture, threat vector, trust, and compliance in their taxonomy. The terminology is not aligned with the one in the field. Trust is according to the authors related to privacy, availability, and reliability, which might be the case for end-users, but not (only) for other entities.  Shasha et al.~\cite{8502818} use simple differentiations in their taxonomy, such as physical, nearby, and remote, not taking all aspects into account.

Other authors establish a taxonomy based on their survey results. Williams et al.~\cite{8884913} conducted a survey to classify security features and threats in IoT devices. The authors use a simple list of seven items, which they call taxonomy. In consequence, they do not include all issues. Similarly, Squillace and Bantan~\cite{9817225} conducted a study with limited results. Xenofontos et al.~\cite{9430606} propose an attack taxonomy for IoT and study different cases of IoT insecurity. Although the case study is extensive, the taxonomy is comparably simple. In consequence, several aspects can be adapted, though none of these IoT attack taxonomies is sufficient to describe attacks on identities related to IoT. Nonetheless, they do not propose a taxonomy.

Several approaches focus on specific aspects. Taivalsaari and Mikkonen~\cite{8354417} propose a taxonomy related to IoT client security, which is one aspect of IoT security. Auliar and Bekaroo~\cite{9590841} focus on IoT security taxonomy related to the Mirai botnet, whereas El-hajj et al.~\cite{8305419} analyze the security of IoT authentication in form of a taxonomy. Lounis and Zulkernine~\cite{9090905} propose a taxonomy related to the security of short-range wireless technology for IoT devices. Boujezza et al.~\cite{7507266} establish a taxonomy related to identity management for the IoT environment, which is already outdated. Although Bikos and Kumar~\cite{9761252} use well-defined layers application, middleware, access gateway, and edge technology, they mainly focus on the usage of blockchain for IoT. Berger et al.~\cite{10.1145/3462513} focus on resilience for IoT devices. Alsubaei et al.~\cite{8110212} propose a taxonomy related to the security of medical IoT devices. Their differentiation is mainly on availability, integrity, and confidentiality as well as a few layers.

\subsection{Attacks on Identities Management}

As shown, different attacks on identity management are possible. Fritsch~\cite{mci/Fritsch2020} identifies identity management as a target in cyberwar. Our previous work~\cite{10.1145/3538969.3544430} proposes a first taxonomy on attacks. First, we study generic cases and then detail specific areas of identity management.

\subsubsection{Case Study of Attacks}

Several authors focus on attacks related to digital identities. Redding et al.~\cite{9854293} analyze the Parler data breach, which used a massive application programming interface (API) scraping of Parler's servers. This was possible as Parler failed to implement authentication on calls made to the platform's API correctly. In a further step, the attackers uncovered credentials due to insufficient security measures. Similarly, Gibson et al.~\cite{9799221} describe the LinkedIn data breach by massive API scraping in 2021. Also here, failed implementation of authentication and authorization for API calls was one of the reasons. Qian et al.~\cite{9850265} analyze the SocialArks data breach resulting from a brute-force web login attack. Elasticsearch does not have enabled authentication by default, which was one of the problems in this incident. In the next step, the attackers got superuser permissions. Ba et al.~\cite{9799087} describe the case of the Canva data breach, where the attacker GnosticPlayers was able to obtain data from 139 million users by credential stuffing and credential cracking.

Other attacks are more sophisticated. Rizkallah et al.~\cite{9817361} focus on the BlueToad case. It is unclear, how the attacker got hold of the unchangeable UDIDs of Apple. Most likely, Apple shares the unencrypted unique device identifiers (UDIDs) with companies or applications, which then store them in their databases. Attackers may be able to steal them if weak security policies are applied. Pitney et al.~\cite{9854268} systematically review the 2021 Microsoft Exchange data breach exploiting four different zero-day vulnerabilities. The attack methodology includes server-side request forgery, deserialization vulnerability, first file write vulnerability, and second file write vulnerability. As patches were not timely installed, this data breach impacted several organizations. Nadjar et al.~\cite{9850246} analyze the case of the multi-vector data breach on Astoria, where confidential user data was exploited by MySQL and PHP-based vulnerabilities in a popular data management tool. After a data import request, the attackers were able to access a PHP file and obtain the admin credentials. This then leads to data exfiltration. Faircloth et al.~\cite{9817175} describe the brute-force attack on T-Mobile leading to subscriber identity module (SIM) hijacking and identity theft. The attacker used a dictionary attack, rainbow table attack, guessing attack, and spidering. Motero et al.~\cite{9501961} utilize a practical survey to describe attacks on Kerberos authentication protocols. The authors analyze overpass the hash, pass the ticket, golden ticket, silver ticket, Kerberoasting, unrestricted delegation attacks, restricted delegation attacks, resource-based restricted delegation attacks, and Kerberos bronze bit attacks. All these attack descriptions are included in TaxIdMA.

\subsubsection{Attacks on Specific Areas}

Other authors focus on specific applications. Naik et al.~\cite{9659929} propose an attack tree for SSI. Anita and Vijayalakshmi~\cite{8944615} and Saad et al.~\cite{9019870} regard blockchain (used among others for SSI) security based on surveys. Al-Khurafi et al.~\cite{7478735} describe the security of web applications based on a survey. The security of web applications is relevant to the security of digital identities. Gaikwad and Ragha~\cite{8389593} focus on the mitigation of attacks on authenticating identities in ad-hoc networks, while Sharma and Singh~\cite{7164831} describe detection techniques related to it. Bahri~\cite{10.1145/3133956.3136066} outlines identity-related threats in online social networks. Based on a survey, Gupta et al.~\cite{7813778} categorize social engineering attacks with a focus on phishing attacks, where information on digital identities is stolen. Qin et al.~\cite{10.1109/WI-IAT.2013.149} address false identity attacks in peer-to-peer networks. Karunanayake et al~\cite{9471821} categorize de-anonymization attacks on the Tor network.  Similarly, Erdin et al.~\cite{7152825} consider finding hidden users based on a survey. Mavoungou et al.~\cite{7547270} provide a survey on threats and attacks on mobile networks, where, for example, devices have identities. Briones et al.~\cite{6618697} show identity theft in a WiFi setting through a case study. Mei et al.~\cite{9750472} publish a survey on advanced persistent threats (APTs), where digital identities are stolen among other things.

Further authors use the information for their proposals. Barona and Mary Anita~\cite{8074287} summarize data breach challenges in cloud computing security including identity management. These comprise data breaches, account or service traffic hijacking, insecure interfaces and APIs, denial of service (DoS), malicious insiders, abuse of cloud services, and shared technology vulnerabilities. Fang et al.~\cite{8686093} analyze data breaches in underground forums, whereas Subramanian et al.~\cite{9432175} propose a model to predict cyber hacking breaches. Information from the specific applications and data breach challenges are considered as input. General attacks on identity management are summarized by Haber and Rolls~\cite{vectors}. Further attacks are described towards blockchains and wallets, see e.\,g. \cite{10.1145/3297280.3297430}, though not for SSI.

\subsection{Threat Information Sharing}

The information about the previously described threats can be shared by 1) a systematic language and 2) a sharing platform. One well-known language is STIX~\cite{stix}, which specifies attack pattern, campaign, course of action, grouping, identity, indicator, infrastructure, intrusion set, location, malware, malware analysis, note, observed data, opinion, report, threat actor, tool, and vulnerability. Identity has the properties name, description, roles, identity\_class, sectors, and contact\_information and might have relationships. In addition, suspicious action at user account objects can be outlined. Thereby, several attacks can be specified. Ussath et al.~\cite{10.1007/978-3-319-32467-8_20} extend STIX to support complex patterns, whereas Vielberth et al.~\cite{Vielberth2019} add human-as-a-Security-Sensor. For identity management, more information might be needed. The Trusted Automated eXchange of Intelligence Information (TAXII)~\cite{taxii} framework is an application layer protocol for the communication of cyber threat information in a simple and scalable manner and it relates to STIX. Open Indicators of Compromise (OpenIOC)~\cite{openioc} apply schemes and specific terms to describe metadata, criteria, and parameters. This shows that the threat information description is not suitable for identity management. Incident Object Description Exchange Format (IODEF)~\cite{iodef} is an incident object description exchange format representing computer security information exchanged between computer security response teams. It is based on \cite{rfc5070,rfc6685,RFC7203}. The format includes incident ID, related activities, detection time, start and end time, report time, description, assessment, method, contact, event data, history, and additional data. Thereby, it can generally be used. A more detailed description format though would help identity management.

Burger et al.~\cite{10.1145/2663876.2663883} analyze ontologies for sharing threats, such as OpenIOC, STIX, and IODEF and propose their own taxonomy. Stillions~\cite{stillions2014dml} describes a detection maturity level model (DML), which was extended by \cite{bromander2016semantic} to present cyber threats. It thereby specifies the attacker's identity, goals, strategy, tactics, techniques, procedures, and tools, as well as traces of the attack execution. Pahlevan et al.~\cite{10.1145/3465481.3470476} extend the TAXII framework for distributed ledger technologies.  Mavroeidis and Bromander~\cite{8240774} analyze taxonomies, ontologies, and standards for cyber threat sharing. The authors conclude that there is no existing ontology, which can be used within cyber threat intelligence as the existing ones mainly lack expressiveness and cover all relevant data and information. Zibak and Simpson~\cite{10.1145/3339252.3340528} explore the benefits and barriers of threat information sharing, while Stojkovski et al.~\cite{10.1145/3485832.3488030} analyze the user experience. Bromander et al.~\cite{10.1145/3458027} propose a new data model for the exchange, whereas Mavroeidis et al.~\cite{9468305} argue to improve current ontologies by commonly agreed-upon controlled vocabulary. This shows that work is still needed. As we notice that detailed information is missing regarding identity management, this conclusion is especially true for identity management.

One possible sharing platform is the Malware Information Sharing Platform (MISP)~\cite{10.1145/2994539.2994542}.  Another open source platform is Open Cyber Threat Intelligence (OpenCTI)~\cite{opencti}, which allows organizations to manage their cyber threat intelligence knowledge and observables. OpenCTI's knowledge schema is based on the STIX standard. The tool can be integrated with others, such as MISP and TheHive. In contrast, The Hive~\cite{thehive} is a security incident response platform, which can make use of MISP.

\subsection{Limitations of Current Approaches}

Although several attack taxonomies and categorizations exist, none focus on identity management. As a result, a taxonomy for attacks related to identities and identity management systems is still missing.  TaxIdMA~\cite{10.1145/3538969.3544430} is a first approach but requires further work. Elements of the related work (taxonomies, case studies of attacks, etc.) can be used as a basis for a holistic taxonomy resp. an improved TaxIdMA. This taxonomy can be enhanced by threat information-sharing language, which needs to be extended for the purpose of identity management.

\section{Methodology}
\label{sec:methodology}

This section describes the methodology used to design the taxonomy framework. First, criteria for evaluation are established before the steps towards TaxIdMA with its previous version, limitations, and the improved version are outlined. Next, design decisions are justified and the naming convention is specified. The glossary defines the terms used in this article. Last but not least, the limitations are summarized.

\subsection{Criteria for Evaluation}

A taxonomy organizes the concepts hierarchically, while each concept includes a short description and further information. Thereby, a taxonomy can help to define and clarify a specific topic~\cite{wendzel}. Before building a new resp. improving a taxonomy or taxonomy framework, criteria have to be determined for judging its merits~\cite{10.1145/185403.185412,601330}. For this article, the following criteria are selected to judge the effectiveness.

\begin{itemize}
\item \textbf{Completeness/Exhaustibility:} All objects are contained in the taxonomy.
\item \textbf{Comprehensiveness:} The taxonomy is understandable for experts in the field. If the taxonomy is understandable for novices in the field, this would be beneficial.
\item \textbf{Well defined:} The terminology is established in the field, meaning there is no confusion as to what is meant.
\item \textbf{Unambiguousness:} The categories are clearly defined, ensuring there exists no confusion.
\item \textbf{Mutual exclusivity:} Categories do not overlap and thereby prevent ambiguities.
\item \textbf{Replicability:} Repeated attempts at classification result in the same taxonomy classes.
\item \textbf{Versatility:} There is a clear process for adding new items and updating the taxonomy.
\end{itemize}

\subsection{Steps towards TaxIdMA}

This section outlines the steps toward the taxonomies and their iterations. First, the iterations of the previous version are summarized, before the limitations are described. This leads to the iterations for the current version. Thereby, the rationale behind the choice of the taxonomies of TaxIdMA is presented.

\subsubsection{Previous Version}

The first and hence previous version of TaxIdMA~\cite{10.1145/3538969.3544430} was generated through a regression manner by the abstraction of knowledge.

\begin{enumerate}
\item First, all related information was gathered in one taxonomy and extended step-by-step while including information from attacks, taxonomies, and other related work. The items were grouped by known categories and found similarities in properties.
\item With the growing complexity, ways to structure it more clearly were explored. This first resulted in two taxonomies: end-users and identity management systems.
\item By regarding both so-designed taxonomies, many similarities were noticed. If these were non-changeable during the attack, they were separated in an attack background. Thereby, the attack background can be used together with arbitrary taxonomies to explain an attack in detail. In addition, the terminology, notion, and structure were aligned.
\item While including further items based on a literature review, the importance of system identities became clearer, adding another taxonomy.
\item The proposed taxonomies were improved by discussions with experts and the application of selected real-world examples.
\item Last but not least, TaxIdMA was evaluated based on real-world examples and a discussion.
\end{enumerate}

Thereby, the first version of TaxIdMA consisted of the taxonomies attack background, system identities, identity management systems, and end-user identities.

\begin{itemize}
 \item \textbf{Attack Background:} The taxonomy on attack background describes the background of all attacks involving identities and is constant during the attack cycle, which can involve several identities. Categories, where the values may vary either depending on the attack type or during the attack cycle, are included in the more specific taxonomies. Thereby, the attack background is used in every description, whereas the further taxonomies depend on the use case.
\item \textbf{System Identities:} During the cyber kill chain, the attacker typically uses several identities, which are described in the related taxonomy. If the attacker utilizes multiple identities, then the taxonomy can be applied to each of these identities.
\item \textbf{Identity Management Systems:} Depending on the motivation of the attacker, gaining access to the identity management system may be one goal as all accounts (human, devices, etc.) are managed there. As a consequence, the categorization of these attacks can be made by the taxonomy of identity management systems. If an organization operates several identity management systems, which are compromised during an attack, for each system the taxonomy should be applied. With the outsourcing of services including identity management, several entities may be involved. As a result, the taxonomy can be applied to all entities.
\item \textbf{End-User Identities:} Whereas gaining access to an identity management system requires additional effort, attacks on end-user identities are usually with less time effort and less financial gain. Due to the scalability, a financial profit can be made. Therefore, another taxonomy describes these attacks.
\end{itemize}

Considering that an attacker may exploit various identities and identity management systems, several up to all specific taxonomies can once or multiple times be applied in a stepwise way. For example, a spear-phishing attack targets an employee. This is possible due to an identity leak for a service the employee typically uses (End-User Identities). With the spear-phishing attack, the employee installs malware, which gives the attackers access to the computer (System/Service Identities). As the identity management systems were not patched recently, the attackers can attack it after some additional steps (Identity Management Systems). In this example, all taxonomies can be utilized to systematically describe the attack. The direction of the description is not pre-defined and can be either from start to end or vice versa. The attack background generally outlines the attack.

\subsubsection{Limitations of the Previous Version}

While the first version of TaxIdMA was also the first step towards a taxonomy framework related to identities, it had several shortcomings.

\begin{itemize}
\item Not all items were unambiguous as social engineering attacks can, for example, apply hardware attacks. As a result, the human element had to be separated. This was done by simplifying the type of attack.
\item The degree of detail varies between the taxonomies and not all information was included in every taxonomy. In consequence, streamlining items and terminology was required.
\item In order to reuse the taxonomies for incident handling, the naming convention needed to be clearly defined.
\item Although the first version was evaluated based on real-world examples, further validation has improved the outcome. In addition, input from further researchers was not actively sought originally.
\item As suggested in its future work section, additional taxonomies related to IoT and SSI might be needed. As SSI is a completely different identity management model, it requires its own taxonomy. Furthermore, IoT devices have limitations, which should be included in the taxonomy. Both applications are used to explain the way how the taxonomies are derived.
\end{itemize}

\subsubsection{Improved Version}

The improved version described in the following sections is built upon the first version and improved in a step-wise way by expert interviews and a literature review.

\begin{enumerate}
\item The first version is used as a basis.
\item By expert interviews and a wider literature review, the given taxonomies are improved and better structured.
\item To comply with STIX, the name system identity is changed to service identity.
\item In order to provide an easier way to reference categories and elements, a naming convention is established.
\item By further discussions and interviews, additional taxonomies are added: IoT devices and the new research direction SSI.
\item Last but not least, TaxIdMA with its taxonomies is evaluated based on expert interviews, the application of real-world examples, and related work.
\end{enumerate}

In consequence, the improved version of TaxIdMA includes the following applications. The methodology for these taxonomy applications is described in the corresponding sections.

\begin{itemize}
\item \textbf{Internet of Things:} IoT is a technology originating from the field of sensor networks. IoT devices can collect, process, and exchange data via a data communication network. In order to identify objects and describe the relationships with owners and other objects, several methods are applied and new ones are proposed, see for example \cite{9254470,9036139,9221144,8984316,9730173}. To keep up with the technological progress and satisfy the diverse options, a new taxonomy is established.
\item \textbf{Self-Sovereign Identities:} Self-sovereign identities are an approach to digital identities that gives individuals control over the information they use to prove who they are to services on the Internet. The research direction obtains momentum with the new version of the electronic IDentification, Authentication, and trust Services (eIDAS) regulation~\cite{10.1145/3543434.3543652,10.1145/3538969.3543817}. As SSI is different from traditional identity management~\cite{9858139}, at least from the entity and layer perspectives, it is best described in its own taxonomy.
\end{itemize}

\subsection{Justification of Taxonomy Design Decisions}

In this section, we give reasons for modifying, disregarding, and applying related work for resp. to TaxIdMA.

\begin{itemize}
\item \textbf{Terminology:} The previous terminology was based on established terms in the field. Although this is relevant for understanding, it does not reflect the possibility to use the taxonomy for threat intelligence sharing. Therefore, especially STIX terminology was taken into account while improving TaxIdMA. One example is the renaming from system identities to service identities.
\item \textbf{Level of Detail:} STIX especially details the attacker, which is adapted to add further information. The same applies to attack types, which are enhanced by Habiba et al.~\cite{Habiba2014}. Windows integrity levels are omitted as they focus only on Windows systems.
\item \textbf{Categories:}  In addition, related work is used to describe attack types. The categories though do not clearly distinguish between attacks with and attacks without social engineering. Social Engineering is one option, though it could be combined with other attack types. As result, current categorizations are dropped. In order to provide information about the device, this item is added to the service identity taxonomy. Here, it is more relevant than in the other taxonomies, where it is combined with the location. With the chosen categories, not all combinations might be possible. Nonetheless, this approach was selected as it may provide more information. For example, unusual combinations are included and information is not redundant.
\item \textbf{Extendibility:} Due to the different natures of SSI and IoT, these applications are added. A key element of taxonomies is extendibility. In order to provide guidance on extendibility, the steps towards these new taxonomies are described in more detail in the corresponding sections.
\end{itemize}

\subsection{Naming Convention}

The items of the taxonomy are enumerated following the convention \textbf{[T].[C].[I].n}.

\begin{itemize}
\item \textbf{T:} Each taxonomy has a unique name with an abbreviation consisting of two resp. three letters: Background (BG), Service Identities (SI), Identity Management Systems (IMS), End-Users (UE), Internet of Things (IoT), Self-Sovereign Identities (SSI), and Web Application (WA).  Thereby, two letters are the standard use case, while known abbreviations are applied.
\item \textbf{C:} The categories also have a one-letter abbreviation: Attacker (A), Target (T), Identity (I), and Attack (K).
\item \textbf{I:} This is followed by another abbreviation for the items: Type (T), Capabilities (C), Identity (I), Permissions (P), Authenticity (A), Delivery (D), Results (R), and Impact (M).
\item \textbf{n:} The leaves are the enumerated. Sub-leaves are added by a dot an additional number. Further leaves are added accordingly if necessary. The only exception is Others (0) for better extendability.
\end{itemize}

For example, to describe the impostor authenticity of the attacker identity in the attack background, the following notion can be used: \emph{BG.I.A.1}. The naming convention is practically shown in the background in Figure~\ref{fig:background}.

The naming of the taxonomy follows a clear structure. First, the type of the category is outlined. Then the category is detailed. The name consists of one word, besides well-established terminologies, such as resource development or privilege escalation. In consequence, the different words can be combined. BG.I.A.1 is can be named \emph{Background Identity Authenticity Impostor}.

\subsection{Glossary}
\label{sec:glossary}

In the following, definitions for the terms used within the taxonomies are given. These definitions are based on~\cite{10.1145/3538969.3544430} and related work.

\begin{itemize}
\item \textbf{Attack:} The use of an exploit by an adversary to take advantage of a weakness with the intent to achieve a negative impact.
\begin{itemize}
\item \emph{Category:} Targeted weakness of identity management.
\item \emph{Delivery:} Way of conveying the attack.
\item \emph{Impact:} Loss or the consequences which are incurring (effects) due to the attack.
\item \emph{Pattern:} Description of the methodology used by the adversaries to exploit weaknesses.
\item \emph{Results:} Direct consequences (final product) of an attack.
\item \emph{Type:} Classification of the attack.
\item \emph{Vector:} Specific path, method, or scenario exploited.
\item \emph{Vulnerability:} Vulnerability resp. vulnerabilities used in the attack.
\end{itemize}
\item \textbf{Attacker:} Someone who explores methods for breaching weaknesses in a computer system or network.
\begin{itemize}
\item \emph{Capabilities:} Expertise or the ability of the attacker to reach the goal.
\item \emph{Type:} Attributes of the attacker.
\end{itemize}
\item \textbf{Identity:} Digital identity used during the attack.
\begin{itemize}
\item \emph{Amount:} Quantity of targeted identities.
\item \emph{Authenticity:} Attribution of the attacker towards the system during the attack.
\item \emph{Completeness:} State or condition of being complete concerning the identity control takeover.
\item \emph{Directness:}  State of direction of targeting.
\item \emph{Lifecycle:} Stage of attack lifecycle, also known as cyber kill chain~\cite{8551383,strom2018mitre}.
\item \emph{Permissions:} Authorization of the overtaken digital identity.
\item \emph{Timeliness:} State and duration of being timely concerning the identity control takeover.
\item \emph{Type:} Type of digital identity used during the attack.
\end{itemize}
\item \textbf{Target:} A goal designated for an attack.
\begin{itemize}
\item \emph{Characteristics:} Characteristics of the target, which have consequences on attack vectors and impact.
\item \emph{Device:} Device of the attacked target.
\item \emph{Domain:} Area of application of the target, similar to the sector, but in a broader sense.
\item \emph{Level:} Target position in the system stack.
\item \emph{Location:} Particular place in the physical space of the target in relation to the attacker.
\item \emph{Identity:} Position of the target related to the attacker.
\item \emph{Sector:} The area of industry, the target is in.
\item \emph{Type:} Characteristics of the target.
\end{itemize}
\item \textbf{Type:} A grouping based on shared characteristics.
\end{itemize}

\subsection{Limitations}

Although this improved TaxIdMA explains the design and iterations of the taxonomies, it uses the terminology of the research area, which requires basic knowledge. As a consequence, even though the structure should be clear, not all items may be known to novices. This article cannot provide further guidance in form of a tutorial. Instead, a web repository would be needed. As identity management is changing, further taxonomies might be needed in the future. Although we evaluate TaxIdMA on a wider basis, not all aspects might be discovered in this version.

\section{TaxIdMA: Taxonomy on Attacks}
\label{sec:TaxIdMA}

This section describes the taxonomy framework TaxIdMA, which consists of taxonomies related to the attack background, service identities, identity management systems, and end-user identities. The background is constant during the attack cycle. In consequence, the taxonomy on the attack background is applied to all attacks and vulnerabilities. The taxonomies on service identities, identity management systems, and end-user identities further detail the attack resp. vulnerability. An attacker typically applies several service identities during the attack lifecycle, which is described in the related taxonomy. For multiple identities, the taxonomy is applied several times. Identity management systems can pose an interesting goal of an attack, as shown by the Solarwinds Orion attack. The related taxonomy can be used for all involved entities in a cross-organizational system. Last but not least, attacks on end-user identities are categorized according to the taxonomy as they are the goal in broader-scale attacks such as phishing or selected attacks, like spear-phishing. The outlined taxonomies can be applied stepwise way.

\subsection{Attack Background}
\label{sec:attackbackground}

\begin{figure}[!htpb]
\centering
\includegraphics[width=0.85\textwidth]{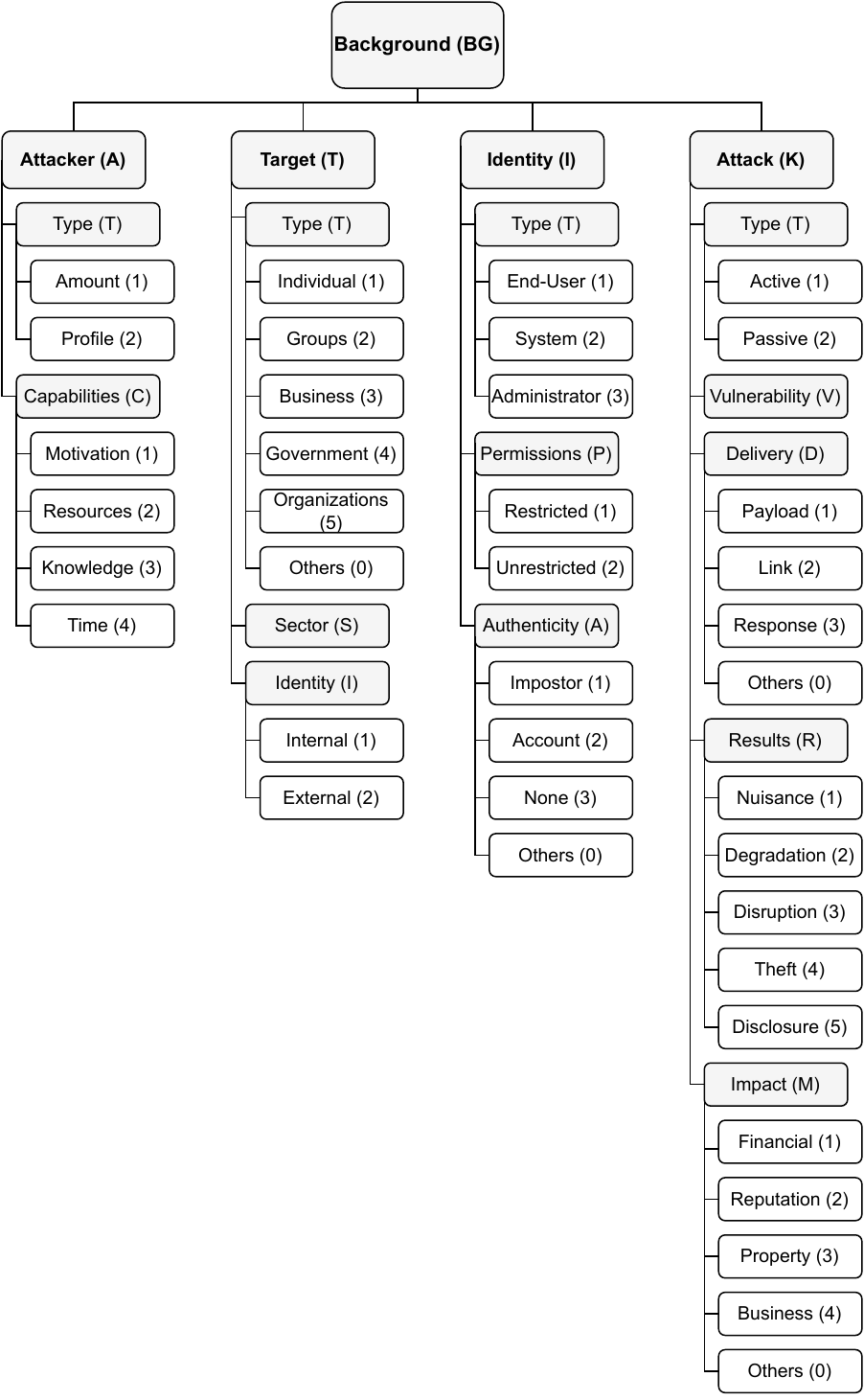}
\caption{Taxonomy for attack background}
\label{fig:background}
\end{figure}

The attack background taxonomy describes the background of the attack, detailed by the following specific attack taxonomies. It is categorized by the attacker, target, attack identity, and the attack itself, as outlined in Figure~\ref{fig:background}.

 \textbf{Attacker:} The attacker is someone who explores methods for breaching weaknesses in a computer system or network. They are detailed by type and capabilities~\cite{CHNG2022100167}.
\begin{itemize}
\item \emph{Type}: The type of attacker describes the position and their profiles.
\begin{itemize}
\item \emph{Amount:} The amount specifies the number of persons involved, ranging from individual to small and big groups resp. organizations.
\item \emph{Profile:} The amount partly relates to the profile. According to STIX, this can be activist, competitor, crime syndicate, criminal, hacker, inside accidental, insider disgruntled, nation-state, sensationalist, spy, terrorist, and unknown. As attacks can be started by script kiddies and other less skilled persons, these should be added.
\end{itemize}
\item \emph{Capabilities}: The expertise or ability of the attacker to reach the goal. The capabilities are characterized by motivation, resources, knowledge, and time. These impact the severity of the attack. The capabilities partly relate to the attacker type.
\begin{itemize}
\item \emph{Motivation:} According to STIX, motivation can be described as accidental, coercion, dominance, ideology, notoriety, organizational gain, personal gain, personal satisfaction, revenge, and unpredictable.
\item \emph{Resources:} Depending on the resources, different attacks are possible. For example, an individual would probably use scripts found online or at Metasploit to exploit, whereas state-sponsored actors might use external sources to implement the malware.
\item \emph{Knowledge:} The sophistication, as STIX calls the knowledge, ranges from none to minimal, intermediate, advanced, expert, innovator, and strategic.
\item \emph{Time:} Time is an important resource, as, for example, scans and brute-force attacks may be extended over a longer time period with the hope of not being noticed by the monitoring system. Therefore, little, medium, and much are possible items.
\end{itemize}
\end{itemize}

\textbf{Target:} A goal designated for an attack, described by identity, type, and sector~\cite{HANSMAN200531,simmons2014avoidit}.
\begin{itemize}
\item \emph{Type:} Attacks focus on different targets, ranging from individuals, groups, businesses, governments, and organizations to other types.
\item \emph{Sector:} These types can be grouped into sectors by applying the notion of STIX.
\item \emph{Identity:} The identity describes the position of the target related to the attacker and can be further detailed by their roles internal (for example, executives, employees, administrators, and contractors) resp. external (for example, partners, customers, trusted third parties, competitors, and strangers).
\end{itemize}

\textbf{Identity:} Identity outlines the digital identity resp. identities used during the attack with their permissions~\cite{10.5555/2048558.2048569} and authenticity.
\begin{itemize}
\item \emph{Type:} The role of the digital identity in use by the attacker, i.\,e., end-user, system, or administrator~\cite{8551383}.
\item \emph{Permissions:} Identities come with permissions according to roles and functions, ranging from restricted to unrestricted~\cite{10.5555/2048558.2048569}. In consequence, permissions describe the authorization the overtaken identity has at the expressed moment.
\item \emph{Authenticity:} The authenticity specifies the authenticity of the attacker towards the system during the attack. The type of identity has one of the following authenticities: impostor (for example, during phishing attacks), the authenticity of a new or compromised account (for example, if successfully attacking a web server or the attacker is able to create a new account), none, or others. The authenticity hence describes one added human element~\cite{8406575}.
\end{itemize}

\textbf{Attack:} The attack is categorized by type, delivery, results, and impact to comply with~\cite{HANSMAN200531,simmons2014avoidit}.
\begin{itemize}
\item \emph{Type:} The type characterizes the threat. Active attacks include social engineering, physical attacks, and web attacks among others. Passive attacks describe eavesdropping and other passive methods.
\item \emph{Vulnerability:} The actual vulnerability exploited by the attacker~\cite{4483667}. This inherently relates to criticality.
\item \emph{Delivery:} This explains the way of delivering the attack, ranging from payloads (for example, a reverse shell), links (for example, phishing links), and responses (for example, server or email responses) to others (for example, physical)~\cite{10.1145/2835375}.
\item \emph{Results:} The direct consequences of an attack, ranging from nuisance and degradation at the lower end to disruption, theft, and disclosure.
\item \emph{Impact:} The loss or the consequences which are incurred due to the attack. This includes financial, reputation, property, business, and others.
\end{itemize}

\subsection{Service Identities}
\label{sec:systemidentities}

\begin{figure}[!htpb]
\centering
\includegraphics[width=0.95\textwidth]{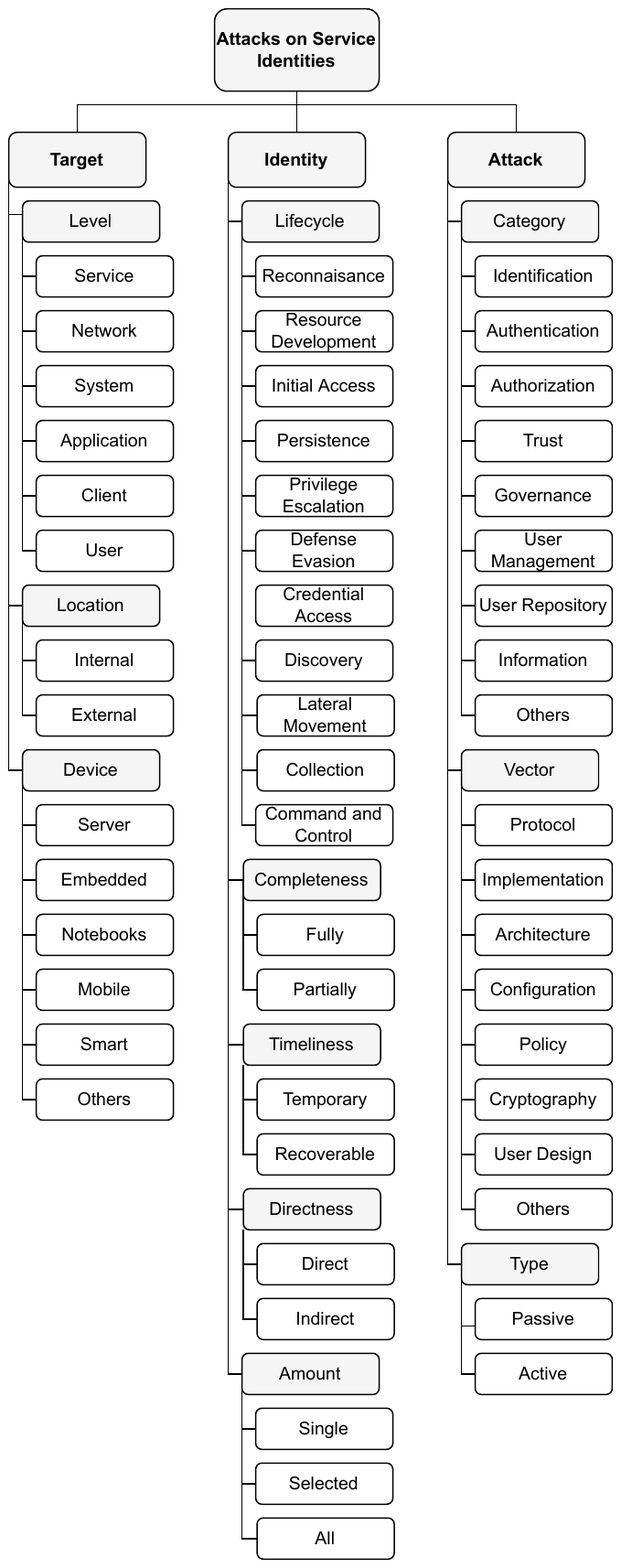}
\caption{Taxonomy for attacks using service identities}
\label{fig:system}
\end{figure}

During attacks targeting servers among others, attackers typically use different identities. In order to categorize these, the following taxonomy (see Figure~\ref{fig:system}) further details target, identity, and attack.

\textbf{Target:} The target specifies the target of the attack. This information is an addition to the attack background.
\begin{itemize}
\item \emph{Level:} Level describes the target level in the system stack. As identities appear on different levels, all these levels can be targeted. This includes a service, network, system with cryptography and hardware, an application with a server (for example, database, storage, web, and email), and a client as well as a user~\cite{bsi}. The degree of detail varies from taxonomy and taxonomy and, therefore, is included in each taxonomy besides background.
\item \emph{Location:} The physical location of the target is categorized here. The location of the target in relation to the attacker may vary, from local/internal to external, for example, a trusted third party~\cite{Habiba2014}.
\item \emph{Device:} The device specifies the location of the targeted device, as it is more relevant in this context.
\end{itemize}

\textbf{Identity:} The identity categorizes lifecycle, completeness, timeliness, directness, and amount.
\begin{itemize}
\item \emph{Lifecycle:} The stage of the attack lifecycle, a.\,k.\,a. cyber kill chain~\cite{8551383,strom2018mitre}.
\item \emph{Completeness:} The completeness of identity takeover, i.\,e., fully or partly.
\item \emph{Timeliness:} The timeliness of identity takeover, i.\,e., definitely temporary or recoverable.
\item \emph{Directness:} The direction of targeting, i.\,e., directly or indirectly.
\item \emph{Amount:} The amount of targeted identities. While applying different identities during the attack lifecycle, the amount is most likely single or selected identities.
\end{itemize}

\textbf{Attack:} The attack is described by category, vector, and type.
\begin{itemize}
\item \emph{Category:} The targeted weakness of identity management, i.\,e., identification, authentication, authorization, trust, governance, user management, user repository, information, or others~\cite{capec}. This includes the identity lifecycle by governance (request, provisioning, and de-provisioning) and identification, authentication, and authorization (operation).
\item \emph{Vector:} The vector specifies the path, method, or scenario exploited. This can range from protocol, implementation, architecture, configuration, policy, and cryptography to user design, and others.
\item \emph{Type:} The type further details the attack, i.\,e., passive or active. Both include further attacks, such as probing, scanning, bypassing, eavesdropping, and modifying~ \cite{8910969}.
\end{itemize}

\subsection{Identity Management Systems}
\label{sec:identitymanagementsystems}

\begin{figure}[!htpb]
\centering
\includegraphics[width=0.95\textwidth]{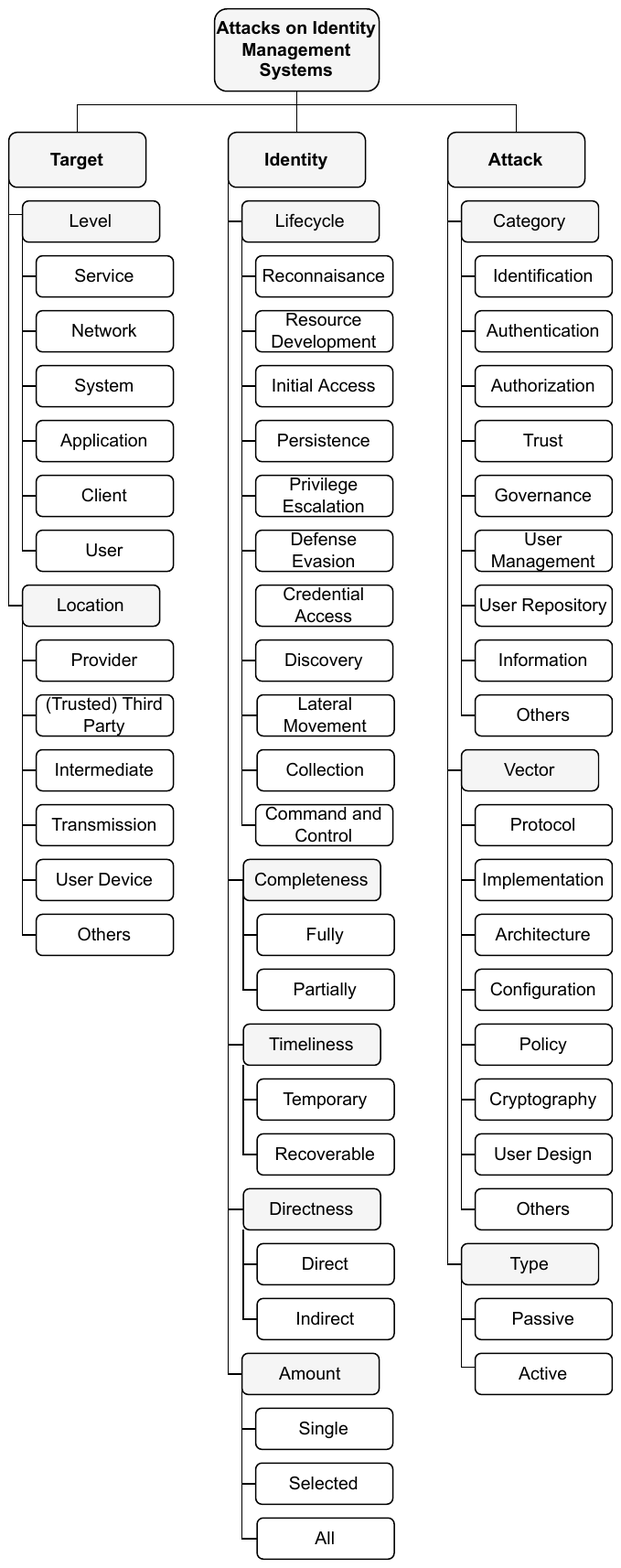}
\caption{Taxonomy for attacks on identity management systems}
\label{fig:idms}
\end{figure}

Due to the reason that identity management systems manage all identities (for example, humans, devices, and services) in an organization, they pose an interesting goal. Figure~\ref{fig:idms} outlines the importance of the location in the diverse setting. Target, identity, and attack are explained in the following.

\textbf{Target:} The target is detailed by level and location.
\begin{itemize}
\item \emph{Level:} The level is similar to the one in service identities~\cite{bsi} shown above and includes service, network, system, application, client, and user.
\item \emph{Location:} Identity management systems can be cross-organizational and operated in a cloud environment by other entities. As a result, the location is either internal or external and according to~\cite{Habiba2014} diverse with (identity/service) provider, (trusted) third party, intermediate, transmission, and user device as well as others. The device is not relevant for identity management systems in contrast to service identities.
\end{itemize}

\textbf{Identity:} The identity is described by the lifecycle, completeness, timeliness, directness, and amount.
\begin{itemize}
\item \emph{Lifecycle:} The attack can involve or target the identity management system at different stages of the lifecycle~\cite{8551383,strom2018mitre}. These include, according to MITRE ATT\&CK, reconnaissance, resource development, initial access, persistence, privilege escalation, defense evasion, credential access, discovery, lateral movement, collection, and command and control.
\item \emph{Completeness:} An identity management system can partly or fully be taken over, as shown with silver and golden tickets for AD~\cite{kerb}.
\item \emph{Timeliness:} The timeliness is either temporary or recoverable, whereas recoverable is standard for an identity management system.
\item \emph{Directness:} The attacker can either directly or indirectly target the identity management system.
\item \emph{Amount:} An attacker can overtake single, selected, or all accounts. The amount might increase during the attack.
\end{itemize}

\textbf{Attack:} An attack is outlined by category, vector, and type. This category is similar to service identities.
\begin{itemize}
\item \emph{Category:} The taxonomy applies the category of attacks by~\cite{capec}; identification, authentication, authorization, trust, governance, user management, user repository, and information as well as others.
\item \emph{Vector:} The attack vector is divided into the items protocol, implementation, architecture, policy, crypto\-graphy, user design, and others. Two examples of implementation are shown next. The implementation of AD had the vulnerabilities MS17-010 Eternal Blue~\cite{eternal} and MS16-032~\cite{logon} in earlier versions. The AD implementation of Kerberos could be used for Pass-the-Hash~\cite{passthehash} and Kerberoasting~\cite{kerberoasting} attacks. The configuration can be a source as well. The configuration of AD has pitfalls, grouped into accounts (for example, password in comments), groups (for example, built-in groups and unlimited groups), and delegation. Some implementations of LDAP are vulnerable to enumeration by misconfiguration. With identity management systems exposed to the web, API security becomes important.
\item \emph{Type:} The type of attack, or technique at MITRE ATT\&CK, describes the sort of attack, which is either passive or active and can be further detailed in the leaves.
\end{itemize}

\subsection{End-User Identities}
\label{sec:enduseridentities}

\begin{figure}[!htpb]
\centering
\includegraphics[width=0.6\textwidth]{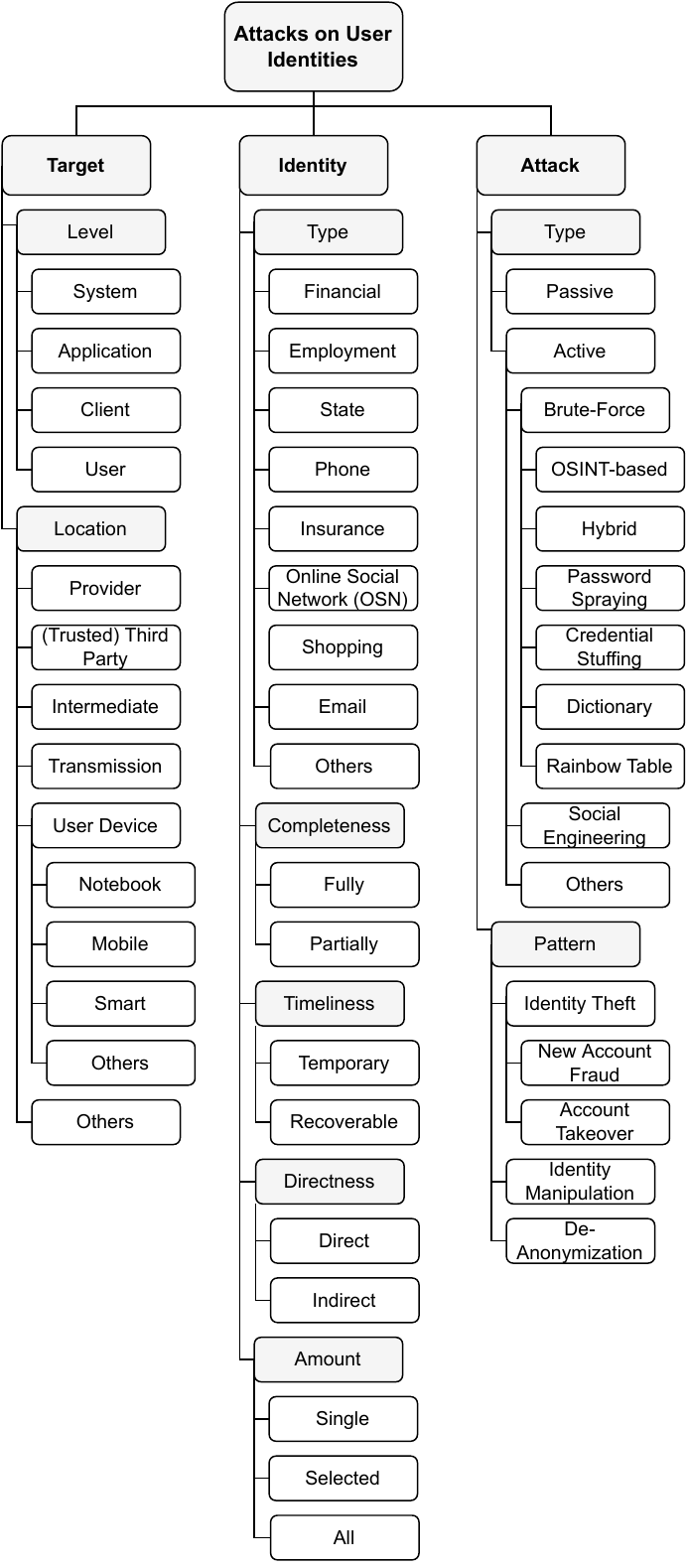}
\caption{Taxonomy for attacks on end-user identities}
\label{fig:identities}
\end{figure}

The end-user identity taxonomy focuses on user identities, which are typically targeted in large-scale attacks. While an individual digital identity has little financial value (which varies between the identity types), the amount of acquired accounts makes these types of attacks interesting for attackers. In consequence, the taxonomy as shown in Figure~\ref{fig:identities} includes additional identity types. The type of attack is concretized by the inclusion of an additional pattern.

\textbf{Target:} The target user limits the possibilities.
\begin{itemize}
\item \emph{Level:} User identities appear on the levels system, application, client, and user.
\item \emph{Location:} User identities are stored in databases and identity management systems mainly. Furthermore, users may store them directly or indirectly on devices. As a result, the same locations are possible with (identity/service) provider, (trusted) third party, intermediate,  transmission, user device, and others.
\end{itemize}

\textbf{Identity:} The identity is described by type, completeness, timeliness, directness, and amount.
\begin{itemize}
\item \emph{Type:} Typical identity types contain information resp. accounts about financial information, such as credit card (including child credit card history) and bank; employment (according to STIX, this may be LDAP, OpenID, remote authentication dial-in user service (RADIUS), UNIX, or Windows local/domain); state-related information such as tax, eID, and social security number; phone; insurance including health care; online social networks (for example, Facebook, Twitter, Skype, and Instagram), online shopping, and other accounts.
\item \emph{Completeness:} Completeness is divided into full (for example, phishing) and partial (for example, session hijacking) take-over.
\item \emph{Timeliness:} Timeliness is defined as either temporary (for example, session hijacking) or recoverable (for example, phishing).
\item \emph{Directness:} The attack can either be direct (for example, phishing) or indirect (for example, supply chain attack).
\item \emph{Amount:} The amount ranges from single to selected and all.
\end{itemize}

\textbf{Attack:} The attack is specified by type and pattern.
\begin{itemize}
\item \emph{Type:} The attack type contains the same categories as the already described taxonomies. Typical attacks towards identities are outlined. This includes brute-force attacks and social engineering, which could be further detailed. Brute-force includes OSINT-based, hybrid, password spraying, credential stuffing, dictionary, and rainbow table. Another type is web attacks with cookie replay and other types of session hijacking among others.
\item \emph{Pattern:} The pattern describes the methodology applied by the adversaries. The attack pattern contains identity theft, identity manipulation, and de-anonymization. Identity theft is further divided into new account fraud (for example, existing profile cloning attack) and account takeover (for example, by account recovery exploit), which can be combined. The category of pattern relates to CAPEC, CWE, and OWASP.
\end{itemize}

\section{Application of TaxIdMA on Specific Areas}

TaxIdMA is a rather generic taxonomy framework related to attacks on identities and identity management systems. Specific areas may have customized properties as indicated by~\cite{10.1145/3538969.3544430}. We outlined two of them, IoT and SSI, which have partly different properties. In consequence, we describe TaxIdMA for IoT and SSI in the following.

\subsection{Internet of Things}

IoT describes physical objects with sensors, processing ability, software, and other technologies that connect and exchange data with other devices and systems. In the consumer market, IoT technology is contained in the concept of smart homes. Otherwise, IoT is used in healthcare systems, industry, and many more.

\subsubsection{Methodology}

In order to classify IoT, we use the search terms \textit{(iot OR "internet of things) AND (taxonomy OR categorization OR classification OR vulnerability)} at IEEE, ACM, USENIX, MDPI, and Springer Link. The results are evaluated in accordance with the search term and then further processed to extract the important aspects. If several approaches contradict, the average proposal is used. In addition, unstructured interviews with experts in the IoT area help to further detail certain characteristics. The thereby extracted characteristics are then compared with the taxonomies of TaxIdMA. If IoT differs in a certain aspect, then the corresponding characteristic is added.

\subsubsection{Application of TaxIdMA on Internet of Things}

IoT architectures consist of IoT devices, maybe gateways, network structures, and central systems for administration and processing that might be in the cloud to allow the IoT devices to communicate with each other. Thereby, IoT architectures consist of different layers. These may depend on the actually applied protocols. Based on~\cite{8110212,9430606}, we summarize them as infrastructure (sensors, gateways, other devices, but also central units), communication (connectivity between the elements, for example, with 5G, WiFi, Bluetooth, low-power wide-area networks (LPWANs)~\cite{9090905}), routing, service (application), a client (for end-users), and others. These layers support IoT devices to collect and process data. With these layers, we also notice various identities: all the devices, users, but also applications, and maybe data. As this architecture goes beyond the ISO/OSI model to include the transformation of data into usable information, the layers are adapted in the taxonomy. In consequence, the object characteristics differ, ranging from automation and intelligence to storage and processing~\cite{8356843}. Due to the different infrastructure, other attacks are possible, such as tag cloning, sensor tracking, rogue access, and tampering~\cite{8110212,9219584,9256294,8688434}. In addition, the attacks at least partly depend on the location (inside vs. outside) and the domain (healthcare, transport, smart homes, robotics, etc.)~\cite{7804660}. The security countermeasures may differ from the type of IoT device. For example, industrial and commercial devices might have more countermeasures than consumer devices, although this is subject to the actual device~\cite{9430606}. The attacks again have various consequences, up to life-threatening situations.

In consequence, the taxonomies on the attack background and service identities are adapted as follows. Again, the background generally describes the attack, whereas the other taxonomies specify the attack in more detail. This includes attacks on the communication layer. The taxonomy for attacks on user identities typically uses the identity type others, as IoT devices are concerned. As shown by Wüstrich et al.~\cite{9219584}, the typical attack vectors apply here. Regarding identity management systems, the identity type (for example, human, device, and application) could be included. Otherwise, these two taxonomies remain the same.

\paragraph{Attack Background}

\begin{figure}[!htpb]
\centering
\includegraphics[width=0.85\textwidth]{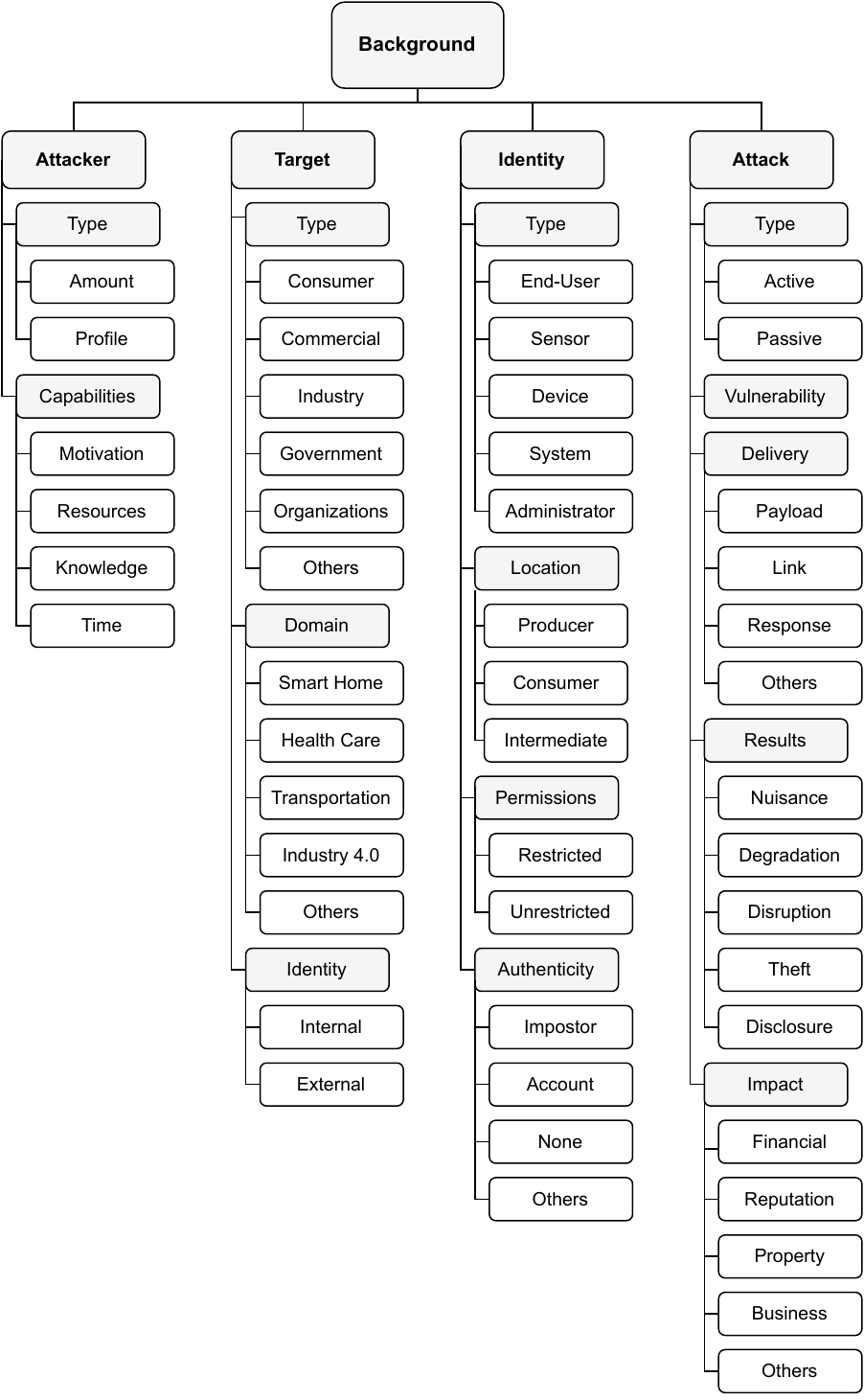}
\caption{Taxonomy for IoT attack background}
\label{fig:background-iot}
\end{figure}

The background, as shown in Figure~\ref{fig:background-iot}, applies the terminology of the IoT area and is adapted as follows.

\begin{description}
\item[Attacker:] Nothing is changed.
\item[Target:] The target type uses the terminology, i.\,e., consumer instead of individual, industry instead of business, and commercial instead of groups \cite{9430606}. Consumer goods target end-user applications including personal devices (for example, cameras, smartphones, and refrigerators). Commercial IoT devices refers to the resources utilized by enterprises and bigger infrastructures. They could be used with an additional security layer for industry, government, and other organizations. Industrial IoT includes sensors, actuators, controllers, industrial assets, remote telemetry, monitoring, and management systems for mission-critical architectures. The sector is changed into a domain, which features the most important areas of smart home, health care, transportation (for example, vehicles), and industry 4.0~\cite{7804660}.
\item[Identity:] The type includes the division into sensor, device, and system. The differentiation between a device and a sensor determines the place of change. As an example, a sensor could be replaced, resulting in different data, while the device stays the same. Furthermore, the locations producer, consumer, and intermediate are added~\cite{9430606}.
\item[Attack:] Nothing is changed.
\end{description}

\paragraph{Service Identity}

\begin{figure}[!htpb]
\centering
\includegraphics[width=\textwidth]{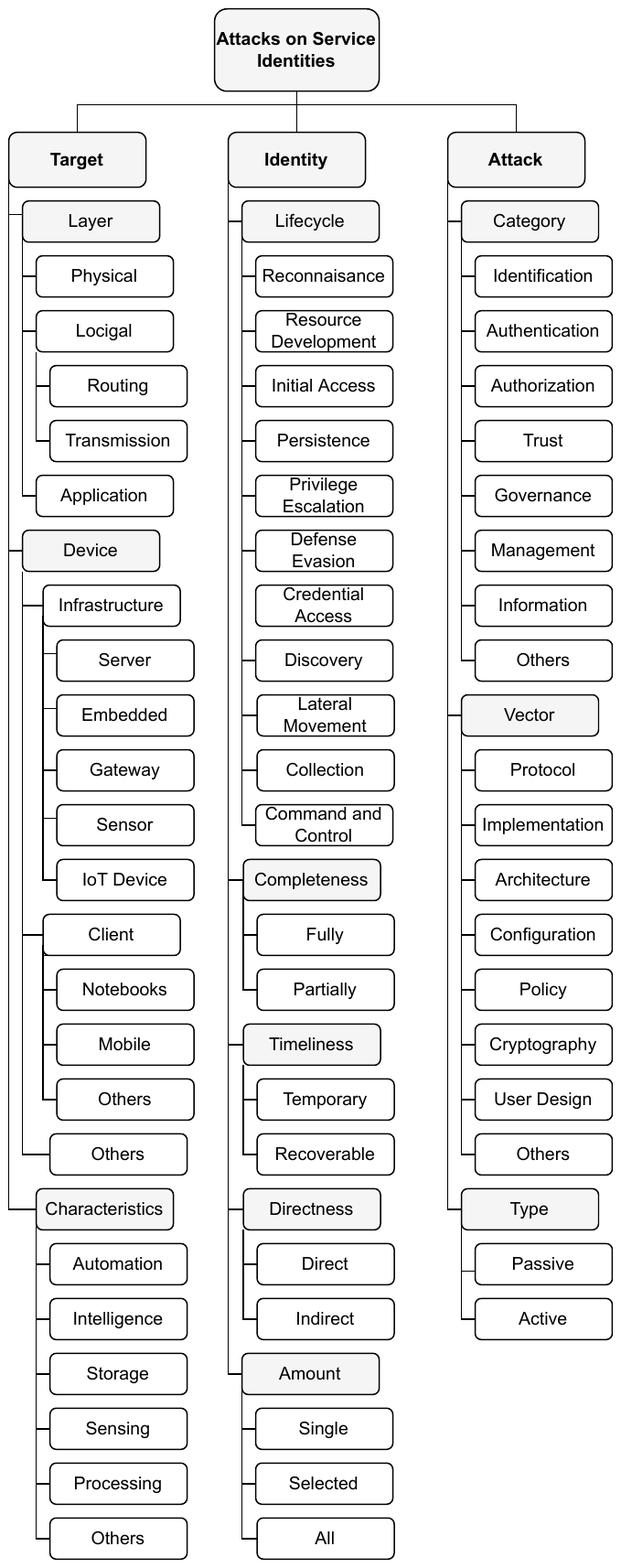}
\caption{Taxonomy for attack IoT service identity}
\label{fig:system-iot}
\end{figure}

Service identities (sensors, gateways, etc.) are adapted according to literature (see Figure~\ref{fig:system-iot}).

\begin{description}
\item[Target:] The target level features the levels (i.\,e., physical, logical, application) summarized above according to the literature~\cite{8110212,9430606}. These can be further detailed if required. Depending on the target level, other attacks are possible. For example, jamming, cloning, and tampering on the physical layer. Exhaustion and unfairness at data link, resp. spoofing, sinkholes, Sybil attacks, wormholes, clone IDs, and more on the network. Applications can be attacked by malware, flooding, and code injection, among others. The location (internal resp. external) is not as important in IoT settings as in others, whereas indoor and outdoor would be more interesting and could be added~\cite{7804660}. The devices reflect the IoT environment with gateways and sensors among others. It is decided to keep both characteristics separately to allow a finer grade of detail. In addition, the characteristics of the target are outlined, i.\,e., automation, intelligence, storage, sensing, processing, and others ~\cite{8356843}.
\item[Identity:] The identity is not changed.
\item[Attack:] The attack category is adapted accordingly, i.\,e., management instead of user management and the omission of user repository. Trust attacks in the IoT environment include, for example, Sybil attacks, bad-mouthing, ballot stuffing, denial of service, black holes, and on-offs~\cite{9315172}.
\end{description}

\subsection{Self-Sovereign Identities}

SSI is an approach that gives users control over their information in a decentralized setting. Thereby, the users can provide user information they received from their home organizations to service providers independently.

\subsubsection{Methodology}

In order to classify SSI, we search for \textit{(attack OR security OR vulnerability) AND (ssi OR "self-sovereign identity" OR "self-sovereign identities)} and elements (wallet, blockchain, distributed ledger technologies) for it at IEEE, ACM, USENIX, MDPI, and Springer Link. We then regard the content for relevance and include the elements accordingly. Based on the literature review and expert interviews, we detail categories and align them in the taxonomy framework accordingly.

\subsubsection{Adaption for SSI}

The SSI architecture comprises the entities issuer, holder, and verifier, which use a decentralized system such as blockchain to store data~\cite{9126742}. Further important components are the wallet of the holder, typically on a smartphone, and software agents and network nodes. The issuer verifies the credentials of the holder, which are then stored in the wallet. The holder can present verifiable presentations depending on the information the verifier requests. Basic information about the entities among others is stored in a decentralized way. This might be blockchain as a form of decentralized ledger technology. One application for blockchain is cryptocurrencies such as Bitcoin, where more attack vectors are known~\cite{9843045,9388487,9825123,9284698,10.1145/3560829.3563557,10.1145/3460120.3484752,10.1145/2976749.2976756,10.1145/3407230,10.1145/3316481}. Consensus and ledger-based attacks are possible, such as Finney attack, race attack, and 51\% attack, resulting in double-spending. If no central authority is introduced, Sybil attacks, eclipse attacks, denial of service attacks, and routing attacks might be possible. Regarding issuer and verifier, fake resp. compromised entities, fake identity attacks, and identity theft attacks are relevant. Another interesting target is the holder with the wallet, which could be compromised by malware, physical access, social engineering, and other forms of malicious actions. More approaches are known for cryptocurrencies~\cite{9071725,8726762,10.1145/3422337.3447832,10.1145/3297280.3297430}.  In consequence, Naik et al.~\cite{9659929} determine software agents, network nodes, and data stores (wallets) as assets. The authors outline faking identity attacks, identity theft attacks, and distributed denial of service attacks as potential attacks on SSI systems. These attacks are then further detailed by attack trees.

Different attacks are possible due to the usage of decentralized storage and protocols. In consequence, the main difference is the target within the taxonomies on end-user identities, identity management systems, and service identities. Especially the items of level and location are adapted (differ from the foundational one) as follows, utilizing standardized terms~\cite{9126742}.

\begin{description}
\item[Level:] Service, Network (i.\,e., normal or decentralized), System (Server, Client), Wallet, Agent, User, Others.
\item[Location:] Issuer, Holder, Verifier, TTP, Decentralized Storage (for example, distributed ledger technologies such as blockchain), User Device, Transmission, Others.
\end{description}

\begin{figure}[!htpb]
\centering
\includegraphics[width=0.6\textwidth]{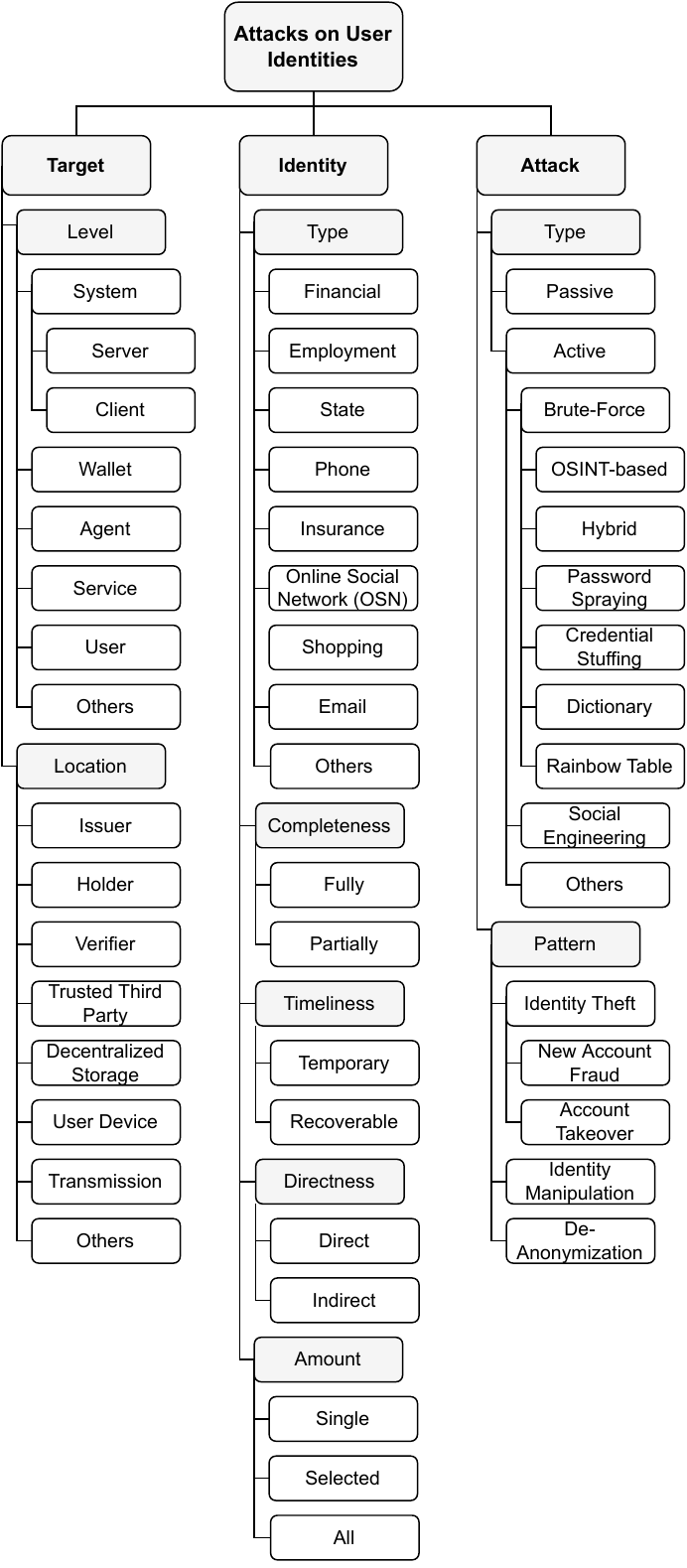}
\caption{Taxonomy for attacks on end-user identities at self-sovereign identities}
\label{fig:identities-ssi}
\end{figure}

The identities do not change as they could be issued to the user. For example, they have the email address user@provider.com. Even though the holder possesses the user device, the importance of the device (typically a smartphone) is emphasized with the addition of both. Due to the fact that SSI gives the user full control over their identities, this is justifiable. Here, especially attacks focusing on the user such as social engineering can be possible. The adapted taxonomy is shown in Figure~\ref{fig:identities-ssi} for end-user identities in the SSI environment.

\section{Evaluation}
\label{sec:evaluation}

The evaluation is three-folded. First, we apply typical threats and talk about statistics concerning identity management. Following this, we evaluate TaxIdMA based on the established requirements. Last but not least, we summarize the expert interviews and their results.

\subsection{Application on Typical Threats}

According to Symantec~\cite{symantec}, targeted attacks are on the rise, whereas 65 \% of groups use spear phishing as the primary infection vector and 96 \% of groups' primary motivation continues to be intelligence gathering. The ENISA threat report 2022~\cite{enisa3} describes that new forms of phishing arise such as spear-phishing, whaling, smishing, and vishing. In consequence, information on the web and social engineering are important ways to receive information. According to the Federal Trade Commission~\cite{ftc}, identity theft is rising since 2001 from 0.33 million to 5.7 million in 2021 in the U.S. Government documents or benefit frauds are on the top with 396,012 complaints, followed by credit card fraud (389,845) and other identity theft (377,203). Further top identity thefts are loan or lease fraud (192,967), bank fraud (124,497), employment or tax-related fraud (111,755), and phone or utilities fraud (88,842) in 2021. Scammers typically engage by phone (646,440), text (378,119), email (264,069), website or apps (180,114), social media (159,458), others (115,730), mail (43,915), resp. online ad or pop-up (36,731). According to Ernst \& Young (EY) Security Survey 2018-19~\cite{ey}, the top ten most valuable information to cyber criminals are customer information (17~\%), financial information (12~\%), strategic plans (12~\%), board member information (11~\%), customer passwords (11~\%), research and development information (9~\%), mergers and acquisitions information (8~\%), intellectual property (6~\%), non-patented IP (5~\%), and supplier information (5~\%). Identity theft is used during all stages of the attack lifecycle. According to ENISA~\cite{enisa,enisa2}, brands such as Microsoft and Amazon are often impersonated. The top data types lost in 2019 are email (65~\%), password (59~\%), name (26~\%), miscellaneous (18~\%), address (13~\%), credit card (12~\%), and account (10~\%). EY~\cite{ey} summarizes the top 10 biggest cyber threats to organizations as phishing (22~\%), malware (20~\%), cyberattacks to disrupt (13~\%), cyberattacks to steal money (12~\%), fraud (10~\%), cyberattacks to steal IP (8~\%), spam (6~\%), internal attacks (5~\%), natural disasters (2~\%), espionage (2~\%).

Regarding the OWASP Top Ten~\cite{topten}, we notice that broken access control is the top web application security risk. This is followed by cryptographic failures, injection, insecure design, security misconfiguration, and vulnerable and outdated components. The seventh item is identification and authentication failures, previously called broken authentication. Software and data integrity failures, security logging and monitoring failures, and server-side request forgery are also included. Thereby, we notice the importance of authentication and access control. This is not only the case at standard online services but also for IoT environments as noticed by~\cite{iotreport}.

In order to take over the credentials of an employee, an attacker might first look at third-party breaches, where actors share datasets publicly or in private. These could be gained from insiders, phished credentials, malware, and otherwise stolen credentials. Based on the dataset, attackers may attempt to compromise the login. If the attacker has a password list, they could try brute-force attempts, password spraying, and password reuse. If the passwords are hashed, then cracking might be successful, if the hash algorithm is insecure and resp. or no salt was added. MFA can be applied to increase security if the factors are independent. Typical methods include email, SMS, and software and hardware code. If the attacker has the original password and MFA is enabled, then further factors need to be bypassed. The bypass attempts depend on the actual deployed MFA and system behind. By finding flaws in the technology underpinning the MFA solution, MFA can be circumvented. This may include compromising the encryption of the secrets, finding patterns, and hijacking name spaces. Again, there is the human element with social engineering. Furthermore, the network session could be hijacked. Email MFA can be bypassed by email (such as O365 mailbox via Exchange Web Services (EWS)) compromise, physical threat, social engineering, and extortion via harassment. For SMS, SIM swapping is an additional attack vector. Social engineering, extortion via harassment, token theft, and physical threat also apply to software and hardware MFA code. Other vectors are the discovery of vulnerabilities by scanning the domain, IP addresses, ports, and services. If a vulnerability is discovered, it might be exploited using an exploit with a customized payload, which is delivered to the victim. Such exploitation results in the initial access to the infrastructure. If the attackers are successful in bypassing the employee's credentials or exploiting the infrastructure, they could access the portal or services and compromise the account. Thereby, further actions such as execution, privilege escalation, etc. might be possible \cite{landscape}. In consequence, the end-user credentials are important here, although other means might be used.

Regarding our taxonomy, we see the following.

\begin{description}
\item[Target:] At least for the first factor, the target level is often the user (i.\,e., social engineering and other password attacks). With MFA, it could not only be user's device, but also the underlying infrastructure and network.
\item[Identity:] The identity type is employment with direct and complete take-over until recovery. The amount can be everything from single to all, although single or selected is most likely.
\item[Attack:] The attack type is most likely active with the pattern of identity theft. If MFA is enabled, then two types are required to compromise the account. The exact type depends on the attack path.
\end{description}

The Identity Defined Security Alliance (IDSA) trend report for 2022~\cite{idsa} outlines that inadequately managed privileges (36~\%), excessive privileges (21~\%), and compromised privileged identities (23~\%) result in breaches. Acquired privileged accounts shorten the attack lifecycle to access sensitive data. Nonetheless, further identities can be used in attacks, especially if the number of identities in organizations is increasing due to the adoption of more cloud applications, third-party relationships, machine identities, and further reasons. Although employees are most likely to be attacked successfully and have the biggest business impact, many organizations have business customers, third parties, consumers, and machine identities. According to IDSA, 84~\% experienced an identity-related breach in the past year. The number of incidents targeting the identity management system is though not known to the authors. At least in the Solarwinds Orion attack, they were used. In consequence, these can be characterized by TaxIdMA.

\subsection{Expert Interviews}

In order to improve and quantitatively evaluate TaxIdMA, we conducted semi-structured interviews with experts in identity management, IoT, and SSI. The interviews were in accordance with the ethics board and their guidelines.

\subsubsection{Methodology}

The experts were selected on the basis of their competence in the fields. Such competence was assumed on the basis of the following main criteria for inclusion: that the experts should be working in the field for at least one, better several years with at least either projects or publications related to their work. For example, a person working in research on the topic of SSI for seven years is considered fitting. The selection of the experts was performed on the authors' personal experience and the advice of other experts. In order to obtain the six experts, we had to contact twelve experts in all. The experts did not get any compensation.

The interviews were performed either in person or virtually via a conference system. The interview language was German (N=5) or English (N=1), according to the preference of the participants. Interviews took place in November resp. December 2022 and lasted an average of 30 minutes. The first part of the semi-structured interviews was dedicated to the specific area of expertise (description of the area and the related threats and attacks). Then, the corresponding taxonomy resp. taxonomies were shown and explained with the text of this article, leading to a discussion about the correctness (general understanding, division of the taxonomies, the related categories and their items, completeness). The third and last part of the interview focused on specific aspects of the taxonomy and the field of expertise (corresponding taxonomies, again threats and attacks, and possible improvements). The transcripts were analyzed, discussed, and collated. Here, we used Delphi panels to discuss and collate the comments.

\subsubsection{Results}

In the following, the results of the expert interviews are summarized.

\paragraph{TaxIdMA:} Before improving TaxIdMA, comments from the presentation of~\cite{10.1145/3538969.3544430} were noted. In addition, a semi-structured interview with two experts was conducted to receive further input for the improved version. The focus was on the correctness, completeness, and understandability of TaxIdMA. Last but not least, typical attack vectors and vulnerabilities were discussed and categorized by TaxIdMA during the interviews.
The feedback was incorporated into the new version of TaxIdMA, which was discussed with five experts (two from the previous round and three new experts) in semi-structured interviews. Here, TaxIdMA was structured and detailed to the satisfaction of the experts and no further iterations were needed.

\paragraph{IoT:} In order to establish a first version of the taxonomy for the IoT environment, related work was analyzed and an expert was interviewed about the potential adaption of TaxIdMA for IoT and different attack vectors and problems with IoT. This was accompanied by another extended literature review, leading to the first version of the taxonomy. The application of the taxonomies was then evaluated by one expert for correctness, completeness, and understandability. While the taxonomy on the background stayed the same, the description was extended by further explanations. The taxonomy of IoT service identities was changed as follows: target layer and device were separated to provide more information and flexibility. This improved taxonomy version was then evaluated by four experts. At this point, the adaption gained the approval of the experts.

\paragraph{SSI:} A first version of the taxonomy was designed based on the literature review and knowledge of the authors. This version of the taxonomy was then discussed with three experts in semi-structured interviews. In the interviews, the experts were fine with the adaptation for SSI. In addition, one expert suggested providing detailed security analyzes of SSI as this is currently missing. Due to the scope of the article, we plan to conduct such an analysis in future work.

\subsection{Requirements}

For TaxIdMA, seven criteria were previously selected to judge the effectiveness. We discuss their fulfillment in the following.

\begin{itemize}
\item \textbf{Completeness/Exhaustibility:} All objects identified by the authors are contained in the taxonomy. There might be objects missed out.
\item \textbf{Comprehensiveness:} As TaxIdMA reuses established terminology and groupings, the taxonomies are understandable for experts in the fields. For novices, further material in form of guidelines and more detailed descriptions would be necessary.
\item \textbf{Well defined:} The terminology is established in the field. In contrast to the original version, some terminology was adapted from STIX, helping to apply TaxIdMA in threat intelligence sharing. As a result, there should be no confusion.
\item \textbf{Unambiguousness:} The categories are clearly defined by the glossary.
\item \textbf{Mutual exclusivity:} In contrast to the original taxonomies, some categories were defined in more detail resp. in less detail, leading to categories not overlapping. Some categories depend though on each other, for example, target level and location. This is needed as further items could be chosen.
\item \textbf{Replicability:} Although different elements of the taxonomies could be grouped differently, the process to derive the so-described taxonomies is outlined, which could result in replicability. With IoT and SSI and the same authors, the same taxonomy classes were used. In consequence, the replicability is at least partly fulfilled.
\item \textbf{Versatility:} The process for updating the original TaxIdMA and adapting the taxonomy framework for IoT and SSI was described in a step-by-step way. Thereby, TaxIdMA is versatile.
\end{itemize}

\section{STIX for Identities}
\label{sec:stix}

STIX is a well-known language and serialization format to exchange cyber threat intelligence related to all aspects of suspicion, compromise, and attribution. It consists of 18 STIX Domain Objects (SDOs), ranging from attack pattern to vulnerability, and two STIX Relationship Objects (SROs), i.\,e., relationship and sighting. Thereby, different attacks are describable in a structured way and this information can be shared. STIX can be extended as long as existing standardized objects or properties are not redefined. The three ways STIX proposes are: 1) define one or more new STIX object types; 2) define additional properties for an existing STIX object type as a nested property extension to represent sub-components or modules; 3) define additional properties for an existing STIX object type at the object's top-level, representing properties that form an inherent part of the definition of an object type. When defining a new STIX object, all common properties associated with that type of object must be included in the schema or definition. In addition, extensions must follow all conformance requirements for that object type. Last but not least, the extension property must be included. TaxIdMA is a taxonomy framework for describing attacks related to identity and identity management. In order to combine both approaches, ways for integration and extension are discussed in this section. The extension should be called \texttt{taxidma v2} with a corresponding ID and will be part of a repository with additional information.

\subsection{Integration of TaxIdMA into STIX}

By the usage of the SDOs attack pattern, campaign, course of action, grouping, identity, indicator, infrastructure, intrusion set, location, malware, malware analysis, note, observed data, opinion, report, threat actor, tool, and vulnerability, the threats are described. Most aspects of TaxIdMA can be incorporated without any problems due to the same or similar terminology. Whereas TaxIdMA concentrates on attacks, the course of action describes countermeasures, which will be future work. With grouping, STIX objects can be shared. Infrastructure further describes the infrastructure of the attackers.  The intrusion set groups the adversarial behaviors and resources with common properties. The location represents the geographical location of either attacker or target and so on. STIX uses attack lifecycles, which are part of TaxIdMA.

\subsection{Extending STIX for TaxIdMA}

Specific SDOs and SROs can be extended by the information specified with TaxIdMA.

\subsubsection{Attack Pattern}

STIX uses \texttt{type} and \texttt{name} (both required) as well as the optional \texttt{external\_reference}, \texttt{description}, \texttt{aliases}, and \texttt{kill\_chain\_phases} as properties. In consequence, the attack pattern relates to malware, identity, location, vulnerability, and tool. A reverse relationship exists with indicator, course of action, campaign, intrusion set, malware, and threat actor.
Regarding TaxIdMA, the kill chain phases can be reused, while TaxIdMA can be given as the external reference. The attack pattern of user identities could extend STIX with \texttt{identity\_pattern}. In addition, \texttt{attack\_type} should be given.

\subsubsection{Campaign}

STIX applies \texttt{type}, \texttt{name} (both required), \texttt{description}, \texttt{aliases}, \texttt{first\_seen}, \texttt{last\_seen}, and \texttt{objective} (all optional). Thereby, the campaign relates to the items intrusion set, threat actor, infrastructure, location, identity, vulnerability, attack pattern, malware, and tool. TaxIdMA uses several to all taxonomies to describe attacks. Thereby, a reference to the different attacks and attack patterns should be included.

\subsubsection{Identity}

STIX requires the properties \texttt{type} and \texttt{name}. In addition, \texttt{description}, \texttt{roles}, \texttt{identity\_class}, \texttt{contact\_information} are optional properties. Identity is located at a location, whereas attack pattern, campaign, intrusion set, malware, threat actor, and tool target identities. Hence, a threat actor may impersonate an identity. Here, a reference to the user account with privileges is missing. The properties \texttt{completeness}, \texttt{timeliness}, \texttt{directness}, \texttt{amount} resp. a list of user accounts, and \texttt{authenticity} can be added.

\subsubsection{Incident}

Incident is currently a stub in STIX 2.1, i.\,e., it is included to support basic use cases but does not contain properties to represent metadata about incidents. Future versions should include these capabilities. Currently, \texttt{type} and \texttt{name} are required if used and \texttt{description} is optional. It could be utilized to combine aspects of the different taxonomies of TaxIdMA.

\subsubsection{Indicator}

The indicator in STIX contains a pattern that can be used to detect suspicious or malicious activities. Required properties are \texttt{type}, \texttt{pattern}, \texttt{pattern\_type}, and \texttt{valid\_from}. Optional properties are \texttt{name}, \texttt{description}, \texttt{indicator\_types}, \texttt{pattern\_version}, \texttt{valid\_until}, and \texttt{kill\_chain\_phases}. The indicator indicates attack pattern, campaign, infrastructure, intrusion set, malware, threat actor, and tool and is based on observed data. The course of action investigates and mitigates indicators. The TaxIdMA \texttt{attack\_category} could be used to further specify the indicator or extend the pattern.

\subsubsection{Intrusion Set}

The STIX intrusion set is a grouped set of adversarial behaviors and resources with common properties. It consists of the required properties \texttt{type} and \texttt{name} and the optional properties \texttt{description}, \texttt{aliases}, \texttt{first\_seen}, \texttt{last\_seen}, \texttt{goals}, \texttt{resource\_level}, \texttt{primary\_motivation}, and \texttt{secondary\_motivation}. The intrusion set is attributed to threat actor, compromises/hosts/owns infrastructure, originates from a location, targets identity, location, and vulnerability, and uses attack pattern, infrastructure, malware, and tool. Regarding TaxIdMA, \texttt{capabilities}, \texttt{impact}, and \texttt{results} could be added.

\subsubsection{STIX Cyber-observable Objects}

The cyber-observable objects describe observations such as artifacts, autonomous systems, directories, email addresses, and more. Related to identities, social engineering and OSINT are important methods for the first steps within the attack lifecycle. In consequence, they are added with \texttt{type}, \texttt{value}, \texttt{description}, and relate to \texttt{identity} and \texttt{location}.

\subsubsection{STIX Vocabulary}

The account type in STIX can have the values \texttt{facebook}, \texttt{ldap}, \texttt{nis}, \texttt{openid}, \texttt{radius}, \texttt{skype}, \texttt{tacacs}, \texttt{twitter}, \texttt{unix}, \texttt{windows-local}, and \texttt{windows-domain}. Further social media accounts, Microsoft, Linux, IoT, mobile devices, etc. are missing and can be added when applying the STIX notation. Infrastructure though uses phishing, but neither other types of social engineering nor other devices such as IoT devices and user devices are currently included.

\subsection{Adding Categories for TaxIdMA to STIX}

In addition, further information systematically collected with TaxIdMA is added to STIX by introducing the following new categories.

\subsubsection{Targeted Organization}

In order to describe the targeted organization when sharing the threat information, \texttt{type}, \texttt{name} (both required), \texttt{sector}, \texttt{domain}, \texttt{description}, and \texttt{size} are applied.

\subsubsection{Device}

To specify the targeted device, \texttt{type}, \texttt{name} (both required), \texttt{level}, \texttt{location}, and \texttt{device\_category} are included.

\subsubsection{Identity Management Category}

If identity management is the goal, a further category can be used. This category further specifies the cyber-observable object software. The properties \texttt{type} and \texttt{name} are required, whereas \texttt{description}, \texttt{vendor}, \texttt{protocol}, \texttt{version}, \texttt{indicator}, \texttt{cpe}, \texttt{swid}, \texttt{languages}, and \texttt{kill\_chain\_phase} are optional.

\section{Discussion}
\label{sec:discussion}

During the design of TaxIdMA and its STIX extension, several iterations were made. These are partly described in the steps towards TaxIdMA. Both TaxIdMA and STIX were discussed with several experts in semistructured interviews.

\subsection{TaxIdMA}

TaxIdMA should fulfill the stated requirements. While the previous version was mostly clear and unambiguous, the current version tries to combat these issues through the outlined changes, which are described in detail in the appendix. Although we discussed TaxIdMA with experts, more experts and real-world attacks could be included to evaluate TaxIdMA extensively. In order to adjust to SSI and IoT, applications of TaxIdMA were designed. Nonetheless, attacks and identity management progress, leading to new changes in the future.

\subsection{STIX}

By extending STIX with TaxIdMA, the systematic description of attacks can be used to share information about the threat with other entities. The extension of STIX was discussed with and improved by experts at the institution. To further evaluate STIX, expert interviews and the application of real-world attacks are necessary. Last but not least, the extension should be tested in an implementation.

\subsection{Open Challenges}

TaxIdMA and the extension of STIX provide a step towards the shift to identities. Further steps are needed. One object of STIX is the course of action, describing countermeasures. Whereas TaxIdMA categorizes attacks, defense mechanisms are still missing. Although the taxonomy for SSI describes several attack vectors, it is not a detailed security analysis, which we plan in future work.

\section{Conclusion and Outlook}
\label{sec:conclusion}

Identities and thereby identity management are important elements of all IT services as everyone and everything has a digital identity for authentication. In consequence, they are essential for IT security. In order to systematically describe attacks and vulnerabilities related to identities and identity management systems, the taxonomy framework TaxIdMA in a revised version was proposed. TaxIdMA consists of four main taxonomies: attack background and the more specific attacks on service identities, end-user identities, and identity management systems. In the improved version, we incorporated input from experts and related work and included a naming convention. In order to improve the previous version, an application on IoT and SSI was introduced. Further adjustments help to clearly specify attacks while keeping the taxonomies as flexible as possible. By describing the iterations towards a taxonomy, future additions are made possible. TaxIdMA is evaluated based on statistics, the application of real-world examples, requirements, and expert interviews. TaxIdMA is accompanied by an extension for STIX to enable the sharing of threat intelligence related to identity management. Thereby, TaxIdMA and STIX work together to increase security. Last but not least, the new version of TaxIdMA is being discussed.

As identity management and attacks resp. threats progress, TaxIdMA will regularly be reevaluated. In future work, we plan to analyze more attacks and common problems and provide a tutorial to better explain TaxIdMA. This tutorial will be published together with STIX in a repository. As shown in the literature review, approaches focussing on the security of SSI are rare. In consequence, we want to analyze threat vectors of SSI in more detail and compare different architectures. Last but not least, defense mechanisms will be grouped in an additional taxonomy framework, which is then mapped to TaxIdMA.

\section*{Acknowledgement}

This article extends~\cite{10.1145/3538969.3544430} as stated in the introduction.

\section*{Conflicts of Interest}

This work is partly funded by the Bavarian Ministry for Digital Affairs (Project DISPUT/STMD-B3-4140-1-4). The authors alone are responsible for the content of the paper.

\section*{Appendix}

\subsection*{Changes to the Taxonomy}

The changes target the following issues:

\begin{itemize}
\item \textbf{Attack Background:}
\begin{itemize}
\item \emph{Attacker:} The position within the attacker type is omitted for clarity as it partly overlapped with the target identity.
\item \emph{Target:} According to literature, the sector is added to the background of the target, while the target type person is changed to individual to suit STIX terminology. The target type group is appended to comply with STIX. Class is though omitted as no benefit was seen.
\item \emph{Identity:} The identity type is changed to comply with typical user levels. In authenticity, temporary is removed as it is nonetheless an account.
\item \emph{Attack:} The attack type is changed to active and passive as although physical, active, passive, offline, and social engineering are typical categories in various taxonomies, they can be combined. For example, a social engineering attack could use physical means. Therefore, the lowest common denominator is chosen. Active attacks though can be further specified. The impact notion is streamlined to fit into one word. In addition, vulnerability is introduced to further detail the background of the attack.
\end{itemize}
\item \textbf{Service Identities:} Name changed from System to Service.
\begin{itemize}
\item \emph{Target:} The location of the target is simplified by only differentiating between internal and external. Furthermore, the corresponding device of the target is added.
\item \emph{Identity:} The item until recovery in timeliness is changed to recoverable to reduce the number of words in this item. The word multiple in amount is modified to all, in order to clearly differentiate to selected.
\item \emph{Attack:} In both, category and pattern, the item others is included. In addition, type is introduced to further specify the attack.
\end{itemize}
\item \textbf{Identity Management Systems:}
\begin{itemize}
\item \emph{Target:} The target locations trusted third party and third party are combined as the difference between them is minimal. In addition, the user is changed to the user's device to further specify the location.
\item \emph{Identity:} The item until recovery in timeliness is modified to recoverable to reduce the number of words in this item. The word multiple in amount is alternated to all, in order to clearly differentiate to selected.
\item \emph{Attack:} In both, category and pattern, the item others is included. In addition, type is introduced to further specify the attack.
\end{itemize}
\item \textbf{User Identities:}
\begin{itemize}
\item \emph{Target:} The target locations trusted third party and third party are merged due to minimal differences. In addition, the user is transformed into the user device to further detail the location.
\item \emph{Identity:} The identity types are rearranged, for example, bank and credit card combined into financial and the missing eID and tax transitioned to state, which could include social security number in the U.S. In addition, the identity type email is added.
\item \emph{Attack:} The attack type is adapted according to the attack background.
\end{itemize}
\end{itemize}

\bibliographystyle{hindawi_bib_style}
\bibliography{exttaxidm}

\end{document}